\documentclass[longauth]{aa} 
\def\kms{km\,s$^{-1}$}

\def\vsini{$v \sin i$}
\def\logg{$\log g$}
\def\Teff{$T_{\rm eff}$}
\def\gdor{$\gamma$\,Dor}
\def\dsct{$\delta$\,Sct}
\def\bcep{$\beta$\,Cep}
\usepackage{graphicx}
\usepackage{longtable,lscape}
\usepackage{txfonts}
\begin{document}

\title{The {\it Kepler} characterization of the variability among A- and F-type stars}
   \subtitle{I. General overview}

   \author{K.~Uytterhoeven
          \inst{1,2,3}
          \and
          A.~Moya\inst{4}
          \and
          A.~Grigahc\`ene\inst{5}
          \and
          J.A.~Guzik\inst{6}
          \and
          J.~Guti\'errez-Soto\inst{7,8,9}
          \and
          B.~Smalley\inst{10}
          \and
          G.~Handler\inst{11,12}
          \and
          L.A.~Balona\inst{13}
          \and
          E.~Niemczura\inst{14}
          \and
          L.~Fox Machado\inst{15}
          \and
          S.~Benatti\inst{16,17}
          \and
           E.~Chapellier\inst{18}
          \and
          A.~Tkachenko\inst{19}
          \and
          R.~Szab\'o\inst{20}
          \and
          J.C.~Su\'arez\inst{7}
          \and
          V.~Ripepi\inst{21}
          \and
          J.~Pascual\inst{7}
          \and
          P.~Mathias\inst{22}   
          \and
          S.~Mart\'{\i}n-Ru\'{\i}z\inst{7}
          \and
          H.~Lehmann\inst{23}
          \and
          J.~Jackiewicz\inst{24}
          \and
          S.~Hekker\inst{25,26}
          \and
          M.~Gruberbauer\inst{27,11}
          \and
          R.A.~Garc\'{\i}a\inst{1}
          \and
          X.~Dumusque\inst{5,28}
          \and
          D.~D\'{\i}az-Fraile\inst{7}
          \and
          P.~Bradley\inst{29}
          \and
          V.~Antoci\inst{11}
          \and
          M.~Roth\inst{2}
          \and
          B.~Leroy\inst{8}
          \and
          S.J.~Murphy\inst{30}
          \and
          P.~De Cat\inst{31}
          \and
          J.~Cuypers\inst{31}
          \and
          H.~Kjeldsen\inst{32}
          \and
          J.~Christensen-Dalsgaard\inst{32}
          \and
          M.~Breger\inst{11,33}
          \and
          A.~Pigulski\inst{14}
          \and
          L.L.~Kiss\inst{20,34}
          \and 
          M.~Still\inst{35}
          \and
          S.E.~Thompson\inst{36}
          \and
          J.~Van Cleve\inst{36}
          }

\offprints{K. Uytterhoeven}

   \institute{Laboratoire AIM, CEA/DSM-CNRS-Universit\'e Paris Diderot; CEA, IRFU, SAp, Centre de Saclay, 91191, Gif-sur-Yvette, France
              \and
Kiepenheuer-Institut f\"ur Sonnenphysik, Sch\"oneckstra\ss{}e 6, 79104 Freiburg im Breisgau, Germany 
\and
Instituto de Astrof\'isica de Canarias, 38200 La Laguna, Tenerife, Spain; Departamento de Astrof\'isica, Universidad de La Laguna, 38205 La Laguna, Tenerife, Spain \\
\email{katrien@iac.es}
\and
              Laboratorio de Astrof\'{\i}sica Estelar y Exoplanetas, LAEX-CAB (INTA-CSIC), PO BOX 78, 28691 Villanueva de la Ca\~nada, Madrid, Spain
             \and
             Centro de Astrof\'{\i}sica, Faculdade de Ci\^encias, Universidade do Porto, Rua das Estrelas, 4150-762 Porto, Portugal 
             \and
             Los Alamos National Laboratory, XTD-2, Los Alamos, NM 87545-2345, USA
             \and
             Instituto de Astrof\'{\i}sica de Andaluc\'{\i}a (CSIC), Apartado 3004, 18080 Granada, Spain
             \and
             LESIA, Observatoire de Paris, CNRS, UPMC, Universit\'e Paris-Diderot, 92195 Meudon, France
             \and
             Valentian International University, Prolongaci\'on C/ Jos\'e Pradas Gallen, s/n 12006 Castell\'on de la Plana - Spain
             \and
             Astrophysics Group, Keele University, Staffordshire, ST5 5BG, United Kingdom
             \and
             Institut f\"ur Astronomie, T\"urkenschanzstra\ss{}e 17, 1180 Wien, Austria
             \and
             Nicolaus Copernicus Astronomical Center, Bartycka 18, 00-716 Warsaw, Poland
             \and
             South African Astronomical Observatory, P.O. Box 9, Observatory 7935, South Africa
             \and
             Instytut Astronomiczny, Uniwersytet Wroc\l{}awski, Kopernika 11, 51-622 Wroc\l{}aw, Poland
             \and
             Observatorio Astron\'omico Nacional, Instituto de Astronom\'{\i}a,  UNAM, Ensenada B.C., Apdo. Postal 877, M\'exico
             \and
             CISAS, Padova University, Via Venezia 15, 35131 Padova, Italy
             \and
             INAF - Astronomical Observatory of Padova, Vicolo Osservatorio 5, 35122 Padova, Italy
             \and
             UMR 6525 H. Fizeau, UNS, CNRS, OCA, Campus Valrose, 06108 Nice Cedex 2, France
              \and
             Instituut voor Sterrenkunde, K.U.Leuven, Celestijnenlaan 200D, 3001 Leuven, Belgium
             \and
             Konkoly Observatory of the Hungarian Academy of Sciences, 1525 Budapest PO Box 67, Hungary
             \and
             INAF - Osservatorio Astronomico di Capodimonte, Via Moiariello 16, 80131 Napoli, Italy
             \and
             Lab. d'Astrophysique de Toulouse-Tarbes, Universit\'e de Toulouse, CNRS, 57 avenue d'Azereix, 65000 Tarbes, France
             \and
             Th\"uringer Landessternwarte Tautenburg, 07778 Tautenburg, Germany
             \and
             Department of Astronomy, New Mexico State University, Las Cruces, NM 88001, USA
             \and
             Astronomical Institute 'Anton Pannekoek', University of Amsterdam, Science Park 904, 1098 XH Amsterdam, The Netherlands 
             \and
             University of Birmingham, School of Physics and Astronomy, Edgbaston, Birmingham B15 2TT, UK
             \and
             Department of Astronomy and Physics, Saint Marys University, Halifax, NS B3H 3C3, Canada
             \and             
             Observatoire de Gen\`eve, Universit\'e de Gen\`eve, 51 ch. des Maillettes, 1290 Sauverny, Switzerland
             \and
             Los Alamos National Laboratory, XCP-6, MS T-087, Los Alamos, NM 87545-2345, USA
             \and
             Jeremiah Horrocks Institute of Astrophysics, University of Central Lancashire, Preston PR1\,2HE, UK
             \and
             Royal Observatory of Belgium, Ringlaan 3, 1180 Brussel, Belgium
             \and
             Department of Physics and Astronomy, University of Aarhus, bygn. 1520, Ny Munkegade, 8000 Aarhus C., Denmark
             \and
             Department of Astronomy, University of Texas, Austin, TX\,78712, USA
             \and
             Sydney Institute for Astronomy, School of Physics, A28, The University of Sydney, NSW, 2006, Australia
             \and
             Bay Area Environmental Research Inst./NASA Ames Research Center, Moffett Field, CA 94035, USA
             \and
             SETI Institute/NASA Ames Research Center, Moffett Field, CA 94035, USA
             }

\date{Received / Accepted}

\abstract{The {\it Kepler} spacecraft is providing time series of photometric data with micromagnitude precision for hundreds of A-F type stars.}{We present a first general characterization of the pulsational behaviour of A-F type stars as observed in the {\it Kepler} light curves of a sample of 750 candidate A-F type stars, and  observationally investigate the relation between $\gamma$\,Doradus (\gdor), $\delta$\,Scuti (\dsct), and hybrid stars.}{We compile a database of physical parameters for the sample stars from the literature and new ground-based observations. We analyse the {\it Kepler} light curve of each star and extract the pulsational frequencies using different frequency analysis methods. We construct two new observables, {\it 'energy'} and {\it 'efficiency'}, related to the driving energy of the pulsation mode and the convective efficiency of the outer convective zone, respectively.}{We propose three main groups to describe the observed variety in pulsating A-F type stars:  \gdor, \dsct, and hybrid stars.  We assign 63\% of our sample to one of the three groups, and identify the remaining part as rotationally modulated/active stars, binaries, stars of different spectral type, or stars that show no clear periodic variability. 23\% of the stars (171 stars) are hybrid stars, which is a much higher fraction than what has been observed before. We characterize for the first time a large number of A-F type stars (475 stars) in terms of number of detected frequencies, frequency range, and typical pulsation amplitudes. The majority of hybrid stars show frequencies with all kinds of periodicities within the \gdor\,and \dsct\,range, also between 5 and 10 d$^{-1}$, which is a challenge for the current models. We find indications for the existence of \dsct\,and \gdor\,stars beyond the edges of the current observational instability strips. The hybrid stars occupy the entire region within the \dsct\,and \gdor\,instability strips and beyond.  Non-variable stars seem to exist within the instability strips. The location of \gdor\,and \dsct\,classes in the (\Teff, \logg)-diagram has been extended. We investigate two newly constructed variables, {\it 'efficiency'} and {\it 'energy'}, as a means to explore the relation between \gdor\,and \dsct\,stars.}{Our results suggest a revision of the current observational instability strips of \dsct\,and \gdor\,stars and imply an investigation of pulsation mechanisms to supplement the $\kappa$\,mechanism and convective blocking effect to drive hybrid pulsations. Accurate physical parameters for all stars are needed to confirm these findings.} 

\maketitle 

\section{Introduction}
With the advent of the asteroseismic space missions MOST (Walker et al. 2003), CoRoT  (Baglin et al. 2006), and {\it Kepler} (Borucki et al. 2010), a new window is opening towards the understanding of the seismic behaviour of A- and F-type pulsators. The main advantages of these space missions are (1) the long-term continuous monitoring of thousands of stars, which enables both the determination of long-period oscillations and the resolving of beat frequencies; and (2) the photometric precision at the level of milli- to micro-magnitudes, which will provide a more complete frequency spectrum and also allow the detection of low-amplitude variations that are unobservable from the ground and providing a more complete frequency spectrum.  The availability of these long-term, very precise light curves makes possible the first comprehensive analysis of the variability of a sample of several hundred candidate A-F type stars that is presented here.

The region of variable A- and F-type, including main sequence (MS), pre-MS, and post-MS stars,  with masses between 1.2 and 2.5 M$_{\odot}$ hosts the $\gamma$\,Doradus (\gdor) and $\delta$\,Scuti (\dsct) pulsators. The \gdor\,stars were recognized as a new class of pulsating stars less than 20 years ago (Balona, Krisciunas \& Cousins 1994).  Our current understanding is that they pulsate in high-order gravity (g) modes (Kaye et al. 1999a), excited by a flux modulation mechanism induced by the upper convective layer (Guzik et al. 2000; Dupret et al. 2004; Grigahc\`ene 2005). Typical \gdor\,periods are between 8~h and 3~d. From the ground, about 70 {\it bona fide} and 88 candidate \gdor\,pulsators have been detected (Balona et al. 1994; Handler 1999; Henry, Fekel \& Henry 2005; De Cat et al. 2006; Henry et al. 2011; among other papers).

The \dsct\,variables, on the other hand, have been known for decades. They show low-order g and pressure (p) modes with periods between 15\,min and 5\,h that are self-excited through the  $\kappa$-mechanism (see reviews by Breger 2000; Handler 2009a). Several hundreds of \dsct\,stars have been observed from the ground (e.g. catalogue by Rodr\'{\i}guez \& Breger 2001).

Because the instability strips of both classes overlap, the existence of hybrid stars, i.e. stars showing pulsations excited by different excitation mechanisms, is expected, and a few candidate hybrid stars have indeed been detected from the ground (Henry \& Fekel 2005;  Uytterhoeven et al. 2008; Handler 2009b).

The main open question in seismic studies of A- and F-type stars concerns the excitation and mode selection mechanism of p and g modes. The only way to understand and find out systematics in the mode-selection mechanism is a determination of pulsation frequencies and pulsation mode parameters for a large number of individual class members for each of
the pulsation classes, and a comparison of the properties of the
different case-studies. So far, a systematic study of a sufficiently substantial sample was hampered by two factors. First, the number of detected well-defined pulsation modes is too small  to construct unique seismic models, which is caused
by ground-based observational constraints, such as bad time-sampling
and a high noise-level. Second, only a small number of
well-studied cases exist, because a proper seismic study requires a long-term
project, involving ground-based multi-site campaigns spanning several seasons, or a dedicated space mission.

First demonstrations of the strength and innovative character of space data with respect to seismic studies of A-F type stars are the detection of two hybrid \gdor-\dsct\,stars by the MOST satellite (HD\,114839, King et al. 2006; BD+18-4914, Rowe et al. 2006), and the detection of an impressive number of frequencies at low amplitudes, including high-degree modes as confirmed by ground-based spectroscopy, in the precise space CoRoT photometry of the \dsct\,stars HD\,50844 (Poretti et al. 2009) and HD\,174936 (Garc\'{\i}a Hern\'andez et al. 2009), and the \gdor\,star HD\,49434 (Chapellier et al. 2011). The first indications that hybrid behaviour might be common in A-F type stars were found from a pilot study of a larger sample of {\it Kepler} and CoRoT stars (Grigahc\`ene et al. 2010; Hareter et al. 2010). Recently, Balona et al. (2011a) announced the detection of \dsct\,and \gdor\,type pulsations in the {\it Kepler} light curves of Ap stars. Hence,  a breakthrough is expected in a currently poorly-understood field of seismic studies of A-F type pulsators through a systematic and careful investigation of the pulsational behaviour in a large sample of stars.

The goals of the current paper are (1) to present a first general characterization of the pulsational behaviour of main-sequence A-F type stars as observed in the {\it Kepler} light curves of a large sample; and (2) to observationally investigate the relation between \gdor\,and \dsct\,stars and the role of hybrids. In forthcoming papers, detailed seismic studies and modelling of selected stars will be presented.  

\section{The {\it Kepler} sample of A-F type stars}
\subsection{The {\it Kepler} data}
\label{sectdata}
The  NASA space mission {\it Kepler} was launched in March 2009 and is designed to search for Earth-size planets in the extended solar neighbourhood (Borucki et al. 2010; Koch et al. 2010). To this end, the spacecraft continuously monitors the brightness of $\sim150\,000$ stars in a fixed area of 105 deg$^2$ in the constellations Cygnus, Lyra,  and Draco, at Galactic latitudes from 6 to 20\,deg. The nearly uninterrupted time series with micromagnitude precision  also opens up opportunities for detailed and in-depth asteroseismic studies with unprecedented precision (Gilliland et al. 2010a).  Of all {\it Kepler} targets, more than 5000 stars have been selected  as potential targets for seismic studies by the {\it Kepler} Asteroseismic Science Consortium, KASC\footnote{http://astro.phys.au.dk/KASC}. 

The {\it Kepler Mission} offers two observing modes: long cadence (LC) and short cadence (SC). The former monitors selected stars with a time resolution of $\sim$30\,min (Jenkins et al. 2010a), the latter provides a 1-minute sampling (Gilliland et al. 2010b). The LC data are well-suited to search for long-period g-mode variations in A-F type stars (periods from a few hours to a few days), while the SC data are needed to unravel the p-mode oscillations (periods of the order of minutes to hours). 

The {\it Kepler} asteroseismic data are  made available to the KASC quarterly. In this paper we consider data from the first year of {\it Kepler} operations: the 9.7\,d Q0 commissioning period (1 May - 11 May 2009), the 33.5\,d Q1 phase data (12 May – 14 June 2009), the 88.9\,d Q2 phase data (19 June - 15 September 2009), the 89.3\,d time string of Q3 (18 September - 16 December 2009), and 89.8\,d of Q4 data (19 December 2009 - 19 March 2010). The SC data are subdivided into three-monthly cycles, labelled, for example, Q3.1, Q3.2 and Q3.3. 

Not all quarters Q0--Q4 are available for all stars. The first year of {\it Kepler} operations was dedicated to the survey phase of the mission. During this phase  as many different stars as possible were monitored with the aim to identify the best potential candidates for seismic studies.  From the survey sample, the KASC working groups selected subsamples of the best seismic candidates for long-term follow-up with {\it Kepler}. From quarter Q5 onwards, only a limited number of selected KASC stars are being observed with {\it Kepler}. The results of the selection process of the most promising \gdor, \dsct, and hybrid candidates are presented in this work. 

\subsection{Selection of the A-F type star sample}
We selected all stars in the {\it Kepler Asteroseismic Science Operations Center} (KASOC) database initially labelled as \gdor\,or \dsct\,candidates. The stars were sorted into these KASOC catagories either because the {\it Kepler Input Catalogue} (KIC; Latham et al. 2005; Brown et al. 2011) value of their effective temperature \Teff\,and gravity \logg\,suggested that they lie in or close to the instability strips of \gdor\,and \dsct\,stars, or because they where proposed as  potential variable A-F type candidates in pre-launch asteroseismic {\it Kepler} observing proposals.  To avoid sampling bias and to aim at completeness of the sample, we analysed all stars listed in the KASOC catalogue as \dsct\,or \gdor\,candidates. Our analysis results provide feedback on the initial guess on variable class assigment by KASOC. As will be seen (Sect.\,\ref{sectotherclass}), several of these stars actually belong to other pulsation classes, many of which are  cool stars. Because there are much fewer B-type stars in the {\it Kepler} field of view than cooler stars, there is a natural selection effect towards cooler stars. We also included stars initially assigned to other pulsation types that showed periodicities typical for \dsct\,and/or \gdor\,stars. We are aware that many more \dsct\,and  \gdor\,candidate stars are being discovered among the KASC targets, but we cannot include all in this study.

\addtocounter{table}{1}

The total sample we considered consists of 750 stars. For 517 stars both LC and SC data are available, while 65 and 168 stars were only observed in SC and LC mode, respectively.  An overview of the A-F type star sample is given in Table~\ref{database}, available in the on-line version of the
paper. The first three columns indicate  the KIC identifier of the star (KIC ID), right ascension (RA), declination (DEC), and {\it Kepler} magnitude (Kp).   The {\it Kepler} bandpass is wider than the typical broad-band filters that are commonly used in optical astronomy (e.g. Johnson $UBVRI$), and can be described as 'white' light. The next three columns provide information on the spectral type (Spectral Type), alternative name of the target (Name), and a comment on its variability (Variable). Information on binarity comes from the {\it The Washington Double Star Catalog} (Worley \& Douglass 1997; Mason et al. 2001), unless  mentioned otherwise. For binary stars labelled with '$\star$', the double star was suspected by inspecting Digitized Sky Survey and 2MASS images by eye. The next set of columns provides information on the {\it Kepler} time series. For each star, the number of datapoints (N datapoints), the total time span of the dataset ($\Delta$T) expressed in d, the longest time gap in the {\it Kepler} light curves ($\delta$T) expressed in d, and the available (range of) quarters in LC (Quarters LC) and SC (Quarters SC) mode are given. 

\subsection{Sample stars in the literature}
Most of the 750 sample stars were previously unstudied. We searched the catalogue by Skiff (2007) and found information on spectral types for only 212 stars. Besides 198 confirmed A- or F-type stars, among which are fourteen chemically peculiar stars, we discovered that stars with a different spectral type also ended up in the sample. There are six known B stars, one M star,  three K stars, and six G-type stars in the sample. The G star KIC\,7548061 (V1154\,Cyg) is a known and well-studied Cepheid (e.g. Pigulski et al. 2009) and is the subject of a dedicated paper based on {\it Kepler} data by Szab\'o et al. (2011). Sixty-two stars are known to belong to multiple systems, including at least fourteen eclipsing binaries (EB; KIC\,1432149, Hartman et al. 2004; KIC\,10206340, Malkov et al. 2006; catalogues by Pr\v{s}a et al. 2011 and Slawson et al. 2011).  Seven stars are only known as '(pulsating) variable stars'. The star KIC\,2987660 (HD\,182634) is reported as  a \dsct\,star by Henry et al. (2001).  Our sample also includes a candidate $\alpha^2$ Canum Venaticorum star, namely KIC\,9851142 or V2094 Cyg (Carrier et al. 2002; Otero 2007). The {\it Kepler} field hosts four open clusters. In our sample at least six known members of NGC\,6819 are included. Also one, eight, and nine members of NGC\,6791, NGC\,6811, and NGC\,6866, respectively, are in our sample. All 750 stars are included in the analysis.

\addtocounter{table}{1}

\section{Physical parameters of the sample stars}
\label{sectphysparam}
Seismic models require accurate values of physical parameters such as \logg, \Teff, metallicity $[M/H]$, and projected rotational velocity \vsini. We compiled an overview of all \Teff, \logg, and \vsini\,values available for the sample stars in Table\,\ref{tableTeff} in the on-line version of the
paper. The different sources include  literature and KIC, along with values derived from new ground-based data. A description of the different sources is given below. The columns of Table\,\ref{tableTeff} are (1) KIC identifier (KIC ID); (2) \Teff\,value from KIC; (3)-(5) \Teff\,values taken from the literature or derived from new ground-based data (Literature); (6) adopted \Teff\,value (Adopted); (7) \logg\,value from KIC; (8)-(9) \logg\,values taken from the literature or derived from new ground-based data (Literature); (10) adopted \logg\,value (Adopted); (11)-(12) \vsini\,values derived from spectroscopic data (Spectra).
Stars that are known to be spectroscopic binaries are flagged $^{\circ}$ behind its KIC identifier (KIC ID). The derived physical parameters of the binary stars have to be considered with caution because the contribution of the binary components might not have been correctly separated. 

\begin{figure}
\centering
\resizebox{0.90\linewidth}{!}{\rotatebox{-90}{\includegraphics{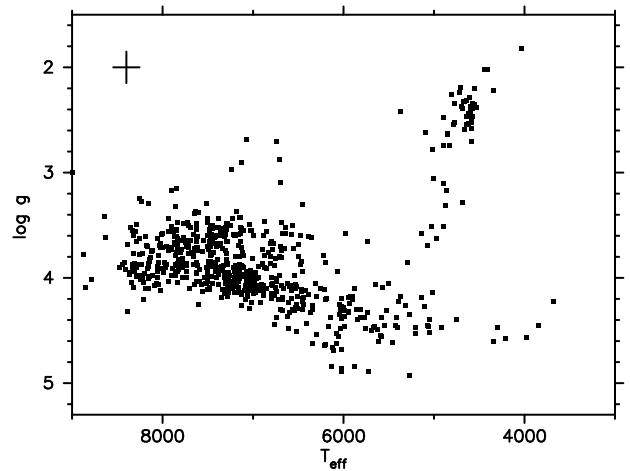}}}
\caption{750 sample stars in the (\Teff, \logg)-diagram. The cross at the left top corner represents the typical error bars on the values:  290\,K for \Teff\,and 0.3\,dex for  \logg.}
\label{Teffloggsample}
\end{figure}

KIC-independent values of  \logg\,and \Teff\,are only available for 110 stars. The values used for the subsequent analysis are  (in order of priority, depending on availability and accuracy) the spectroscopically derived values, or the most recent photometrically derived values. For all other stars we use the only source: the KIC values. The corresponding adopted \Teff\,(in K), and \logg\,(in dex) values are given in boldface in the sixth and tenth column of Table\,\ref{tableTeff}, respectively (column 'Adopted'). For 65 and 71 stars no value of \Teff\,and \logg, respectively, is available. Figure\,\ref{Teffloggsample} shows the sample of 750 stars in the (\Teff, \logg)-diagram.  We estimated the error bars on the KIC values by comparing them with the adopted values taken from the literature or ground-based data. The average difference was 290\,K for \Teff\,and 0.3\,dex for  \logg.  \vsini~values are only available for 52 of the sample stars.

\subsection{Literature}
Besides papers dedicated to specific targets of our sample, the  on-line catalogues by Soubiran et al. (2010),  Lafrasse et al. (2010), Kharchenko et al. (2009), Masana, Jordi,  \& Ribas (2006), Nordstr\"om et al. (2004), Glebocki \& Stawikowski (2000), Allende Prieto \& Lambert (1999), and Wright et al. (2003) were very helpful in the search for values of \Teff, \logg, and \vsini. Also, photometric indices by Hauck \& Mermilliod (1998) were used to estimate values of \Teff\,and \logg. We took care not to include \Teff\,values that are derived from the spectral type rather than directly from data. The literature values of \Teff\,and \logg\,can be found in columns 3--5 and columns 8--9 of Table\,\ref{tableTeff}, respectively ('lit').  We note that the given errors on \Teff\,and \logg, which sometimes seem unrealistic small, are taken from the quoted paper and are not rounded to the number of significant digits. Values of \vsini, expressed in \kms, are given in the last two columns of Table\,\ref{tableTeff}. The source of each value is indicated by the label.

\subsection{{\it Kepler} Input Catalogue}
The KIC provides an estimate of  \Teff\,and  \logg\,for most {\it Kepler} targets derived from {\it Sloan} photometry (see the second and seventh column, 'KIC', of Table\,\ref{tableTeff}, respectively).  Unfortunately, the KIC values of \logg\,are known to have large error bars (Molenda-\.{Z}akowicz et al. 2011, Lehmann et al. 2011). Moreover, a comparison between KIC estimates of the stellar radius and the radius derived from evolutionary models indicate that the KIC values  of \logg\,might be shifted towards lower values by about 0.1\,dex. The temperature values, on the other hand, are fairly good for A-F type stars, and become less reliable for more  massive or peculiar stars, because for higher temperatures the interstellar reddening is apparently not properly taken into account. The stars in our sample are reddened up to 0.3\,mag in $(B-V)$, with an average reddening of $E(B-V) = 0.04$\,mag. The 85 stars for which no KIC \Teff\,value is available, which are generally faint stars (Kp\,$> 11$\,mag), are not considered in any analysis related to temperature, unless values of \Teff\,exist in the literature or are available from the analysis of new ground-based observations (see below).

\subsection{New ground-based observations of sample stars}
\label{sectGB}
In the framework of the ground-based observational project for the characterization of KASC targets (see Uytterhoeven et al. 2010a,b for an overview), targets of the A-F type sample are being observed using multi-colour photometry and/or
high-to-mid-resolution spectroscopy. The goal is to obtain precise values of physical parameters that are needed for the seismic modelling of the stars. A detailed analysis of a first subsample of A-F type stars has been presented by Catanzaro et al. (2011). Several other papers are in preparation. We include the available results to date in this paper, because the precise values of \Teff\,and \logg\,are needed for the interpretations in Sections\,\ref{Sectcharacterization} and  \ref{Sectenergy}.

\subsubsection{Str\"omgren photometry from the Observatorio San Pedro M\'artir}
Multi-colour observations were obtained for 48 sample stars over the period 2010 June 13--17 with the six-channel $uvby-\beta$ Str\"omgren spectrophotometer attached to the 1.5-m telescope at the Observatorio Astr\'onomico Nacional-San Pedro M\'artir (OAN-SPM), Baja California, Mexico.  Each night, a set of standard stars was  observed  to transform instrumental observations into the standard system using the
well known transformation relations given by Str\"omgren (1966), and to correct for atmospheric extinction.  Next, the photometric data were de-reddeded using Moon's {\sc UVBYBETA} programme (Moon 1985), and \Teff\,and \logg\,values were obtained using the $uvby$ grid presented by Smalley \& Kupka (1997). A detailed description of the data will be given by Fox Machado et al. (in prep.).  The resulting stellar atmospheric parameters are presented in Table\,\ref{tableTeff} under label '$b$'.

\subsubsection{SOPHIE spectra from the Observatoire de Haute Provence}
We also analysed spectra of two sample stars, KIC\,11253226 and KIC\,11447883,  obtained during the nights of 2009 July 31, August 1, and August 5 with the high-resolution (R$\sim$70000) spectrograph SOPHIE, which is attached to the 1.93-m telescope at the Observatoire de Haute Provence (OHP), France. The spectra were reduced using a software package directly adapted from HARPS, subsequently corrected to the heliocentric frame, and manually normalized  by fitting a cubic spline.

To derive stellar atmospheric parameters, the observed spectra,  which covers the wavelength range 3870--6940 \AA, were compared with synthetic spectra. The synthetic spectra were computed with the SYNTHE code (Kurucz 1993), using atmospheric models computed with the line-blanketed LTE ATLAS9 code (Kurucz 1993). The parameters were derived using the methodology presented in Niemczura, Morel \& Aerts (2009) which relies on an efficient spectral synthesis based on a least-squares optimisation algorithm.  The resulting values of \Teff, \logg\,and \vsini\,are presented in Table\,\ref{tableTeff}, under label '$h$'. The detailed analysis results, including element abundances and microturbulence, will be presented in a dedicated paper (Niemczura et al. in prep.), including several other {\it Kepler} stars. 

\subsubsection{Spectra from the Tautenburg Observatory}

Spectra of 26 sample stars were obtained from May to August 2010 with the
Coude-\'Echelle spectrograph attached to the 2-m telescope of the Th\"uringer
Landessternwarte Tautenburg (TLS), Germany. The spectra cover 4700 to 7400\,\AA\ in wavelength range,
with a resolution of R $= 32\,000$. The spectra were reduced using standard ESO-MIDAS
packages. We obtained between two and seven spectra per star, which were radial
velocity corrected and co-added. The resulting signal-to-noise in the continua
is between 150 and 250.

Stellar parameters such as \Teff, \logg, [M/H], and \vsini~have been determined by a 
comparison of the observed spectra with synthetic ones, where we used the 
spectral range 4740 to 5800\,\AA, which is almost free of telluric contributions. The synthetic spectra have been computed with the SynthV programme (Tsymbal 1996)
based on atmosphere models computed with LLmodels (Shulyak et al. 2004). Scaled solar abundances have been used for different values of [M/H]. A
detailed description of the applied method can be found in Lehmann et al.
(2011).  The resulting values of \Teff, \logg\,and \vsini~are presented in Table\,\ref{tableTeff}, under label '$g$'. Errors are determined from $\chi^2$ statistics and represent a 1-$\sigma$ confidence level. Detailed analysis results, including also values of [M/H] and microturbulent velocity, will be published in a dedicated paper (Tkachenko et al. in prep.).

\section{Characterization of the sample}
\begin{figure}
\centering
\begin{tabular}{cc}\setlength{\tabcolsep}{0.1pt}
\hspace{-4mm}\resizebox{0.49\linewidth}{!}{\includegraphics{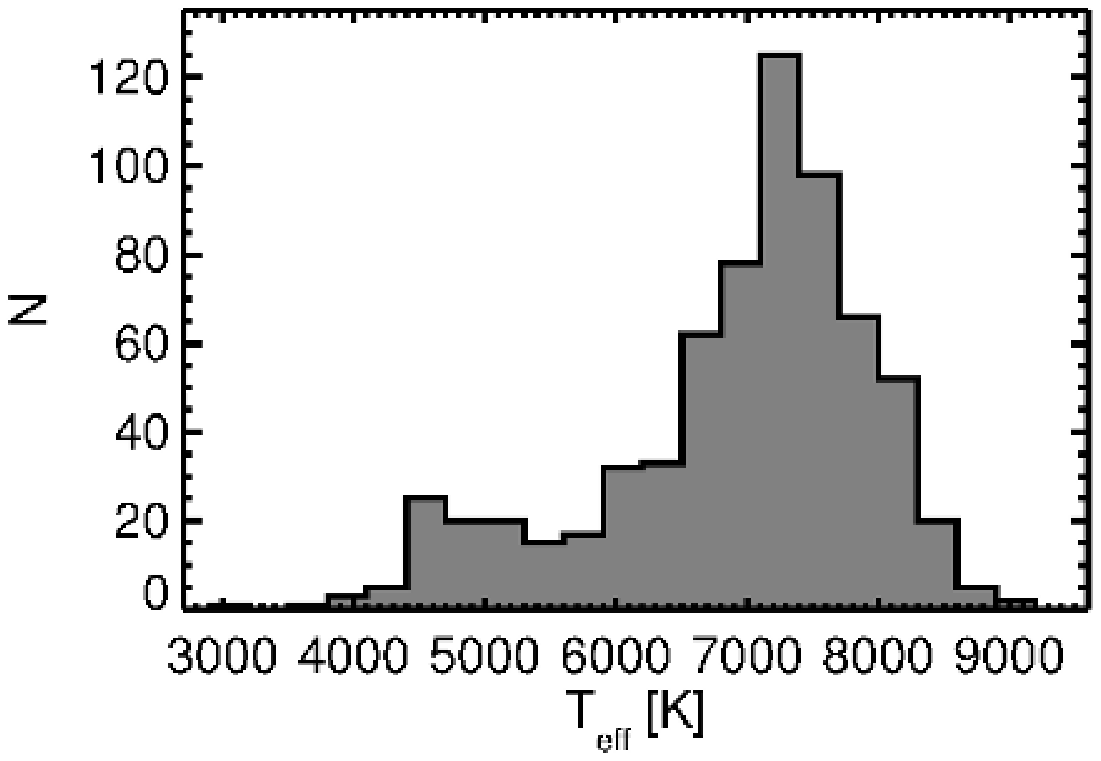}}&
\hspace{-4mm}\resizebox{0.49\linewidth}{!}{\includegraphics{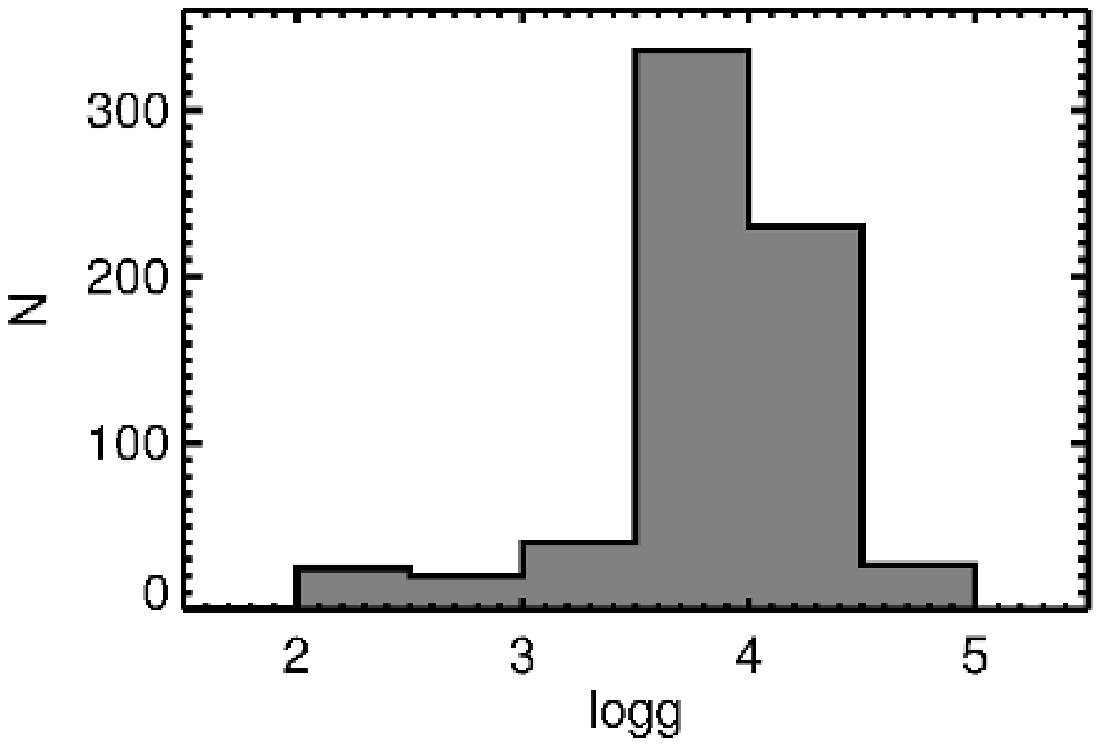}}\\
\hspace{-4mm}\resizebox{0.49\linewidth}{!}{\includegraphics{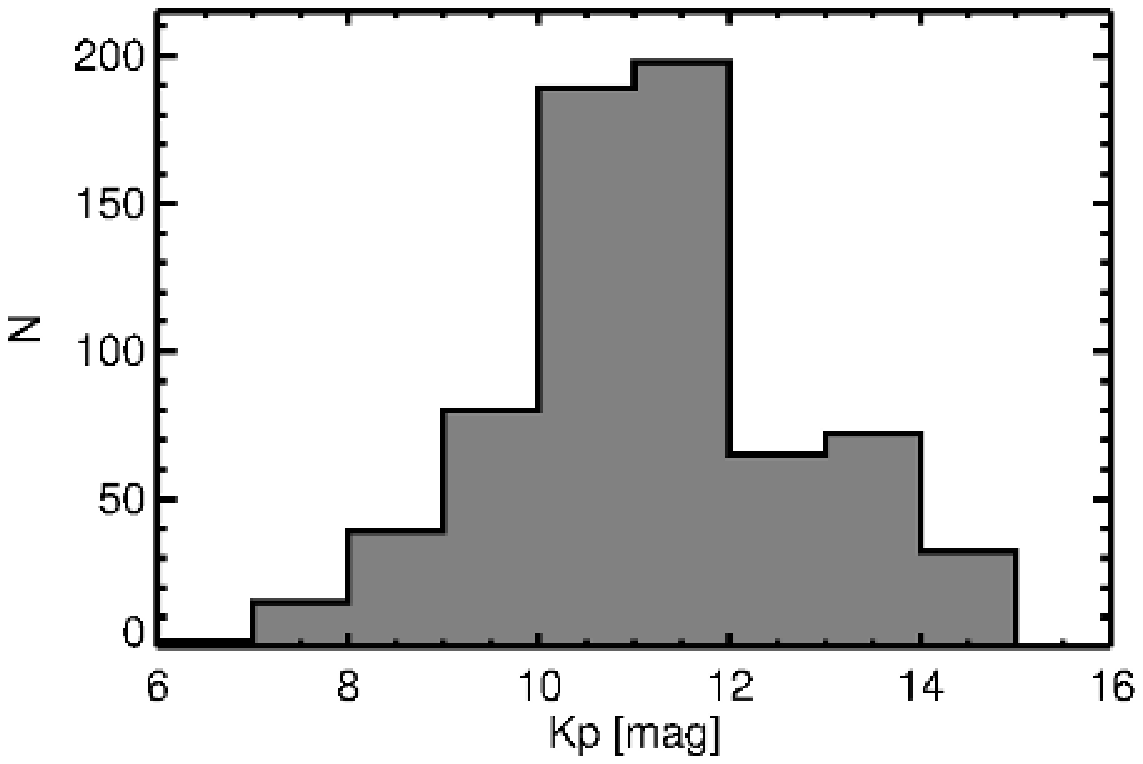}}&
\hspace{-4mm}\resizebox{0.49\linewidth}{!}{\includegraphics{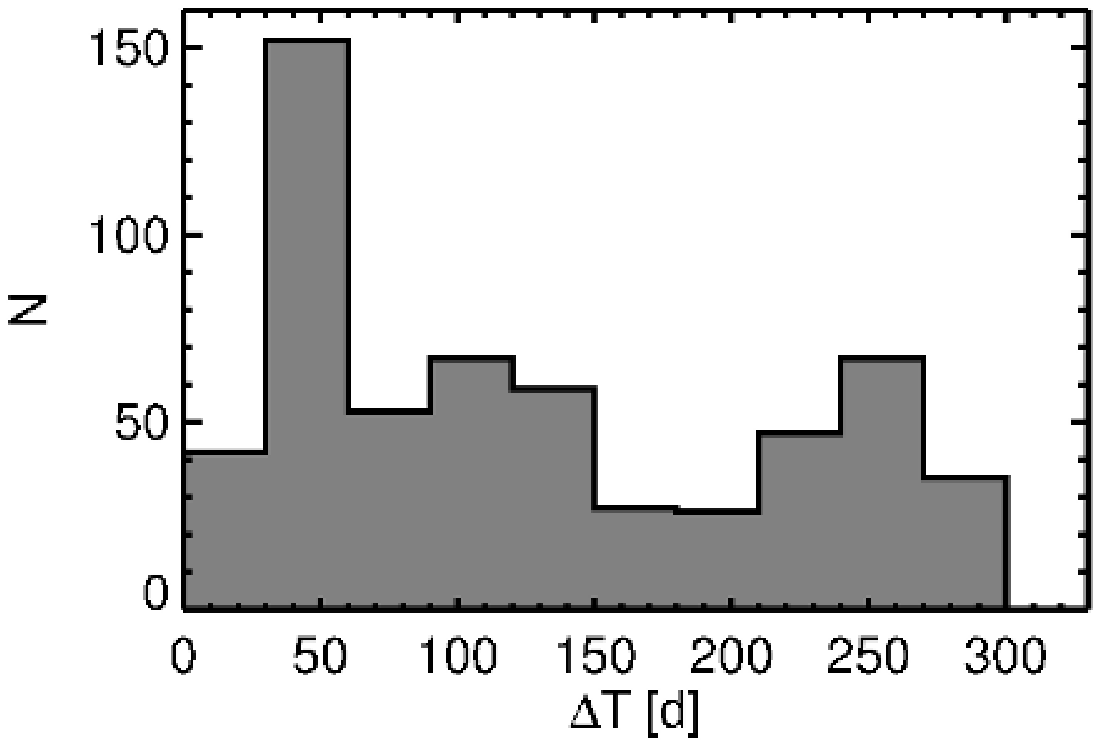}}\\
\end{tabular}
\caption{Distribution  in \Teff\,(top left), \logg\,(top right), {\it Kepler} magnitude Kp (bottom left), and total time span $\Delta$T of the {\it Kepler} light curves, expressed in d (bottom right) of the 750 sample stars. The number of stars belonging to each bin (N) is indicated on the Y-axis. We used the adopted values of \Teff\,and \logg, as given in boldface in Table\,\ref{tableTeff}.}
\label{750stars}
\end{figure}

Figure\,\ref{750stars} shows the distribution of the 750 sample stars in \Teff\,(top left), \logg\,(top right), {\it Kepler} magnitude Kp (bottom left), and total length of the {\it Kepler} light curve $\Delta$T, expressed in d (bottom right). For the analysis we used \Teff\,and \logg\,values given in boldface in Table\,\ref{tableTeff}. Note that seven stars in our sample are hotter than $T_{\rm eff} = 9000$\,K, and fall off the diagram. 

The following typical global parameters have been observed for  \dsct\,and \gdor\,stars (e.g. Rodr\'{\i}guez \& Breger 2001; Handler \& Shobbrook 2002):  \logg $= 3.2-4.3$ and \Teff $= 6300-8600$\,K for \dsct\,stars, and \logg$=3.9-4.3$ and \Teff $= 6900-7500$\,K for \gdor\,stars. While \gdor\,stars are generally MS stars, several more evolved \dsct\,stars have been observed.

The distributions in Fig.\,\ref{750stars} show that about 70\% of the total sample does indeed have \Teff\,values between 6300\,K and 8600\,K. However, a significant number (about 20\%) are cooler stars. The \logg\,values of our sample are concentrated on $3.5-4.5$, which represents about 76\% of the total sample.

The sample consists of stars with magnitudes in the range $6<$\,Kp\,$<15$\,mag. The majority (about 55\%) is located in the interval Kp\,$= [10,12]$\,mag. Given that stars with magnitudes fainter than $V = 9$ are difficult to monitor spectroscopically from the ground with 2\,m-class telescopes, the fact that about 92\% of the stars are fainter than $V = 9$ has implications for the feasibility of possible spectroscopic ground-based follow-up observations (see Uytterhoeven et al. 2010a,b). 

Finally, the total length of the {\it Kepler} dataset (not taking into account possible gaps of several tens of days) is spread between 9.5 and 322~d. For a considerable fraction (19\%) of the sample only Q0 and Q1 data are available, with a total length of 44~d,  implying a frequency resolution slightly worse than 0.02\,d$^{-1}$. On the other hand, 351 stars, or 47\% of the total sample, have a time span of more than 200~d (resulting in a frequency resolution better than 0.005\,d$^{-1}$). Of these 351 stars, 46\% have a maximum time gap in the light curve of less than 10~d, and 23\% have a gap of over 200~d and up to 325~d. 

In the following sections we will describe the variability analysis results of all 750 stars. At this stage we did not exclude any of the stars from the sample on grounds of non-compatibility of physical parameters with the current expectations for A- and F-type pulsators, to present a homogenous analysis and to investigate if  {\it Kepler} confirms the current understanding of \dsct\,and \gdor\,stars.

\section{Frequency analysis}
\subsection{Treatment of the {\it Kepler} light curves}
\label{secttreatment}
 In this paper we used the 'non-corrected' light curves available to KASC for asteroseismic investigations through the KASOC database. A description of the {\it Kepler} data reduction pipeline is given by Jenkins et al. (2010a,b). However, these raw time series suffer from some instrumental perturbations that need to be corrected for, e.g. perturbations caused by the heating and cooling down of the {\it Kepler} CCDs, variations caused by changes in the aperture size of the source mask, etc.  Some of the effects are well known, and the corresponding non-stellar frequencies are tabulated  by the {\it Kepler} team (e.g. frequencies near 32, 400, 430, and 690\,d$^{-1}$). Other perturbations are not documented, and are harder to evaluate and correct for.

We subjected the light curves of all sample stars to an automated procedure that involves fitting a cubic spline to the time series, and correcting the residuals for discontinuities and outliers. To investigate if and to what extent artificial periodicities at the same timescale as the expected pulsations in \gdor\,and \dsct\,stars are introduced by the correction, we also corrected a subsample of stars by a different procedure that takes three types of effects into account, namely outliers, jumps, and drifts (see Garc\'{\i}a et al. 2011). Both correction methods gave the same frequency analysis results within the accuracy of the dataset.

Next, the {\it Kepler} flux ($F_{\rm Kp}(t)$) was converted to parts-per-million (ppm) ($F_{\rm ppm}(t)$) , using the following formula:
\begin{equation}
F_{\rm ppm}(t) = 10^6 \times \left(\frac{F_{\rm Kp}(t)}{\rm f(t)} - 1\right), 
\end{equation}
with {\it f(t)} a polynomial fit to the light curve. A test on the effect of the use of different polynomial orders (2 to 10)  on the detected frequencies in the time series showed that, in general, a third or fourth order polynomial fits the overall curvature better than a linear fit. The choice of the polynomial did not change periodicities with frequencies higher than 0.2\,d$^{-1}$.  The obtained error for frequencies between 0.01 and 0.2\,d$^{-1}$ was of the order of 1/$\Delta$T \,d$^{-1}$, with $\Delta$T the total time span of the light curve expressed in d. 

\subsection{Frequency analysis}

The {\it Kepler} time series of the 750 sample stars were analysed in a homogenous way, using a programme based on the Lomb-Scargle analysis method (Scargle 1982).  Frequencies were extracted in an iterative way until the  Scargle false alarm probability (fap; Scargle 1982), a measure for the significance of a peak with respect to the underlying noise level, reached 0.001. In view of the almost uninterrupted and equidistant sampling of the {\it Kepler} data, this estimate of the fap is a fast and reliable approximation of the true fap, because the number of independent frequencies can be estimated precisely (see also the discussion in Sect.\,4 of Balona et al. 2011b). Frequencies were calculated with an oversampling factor of 10. Time series consisting of only LC data were not searched for periods shorter than 1 hour, because the corresponding Nyquist frequency is 24 \,d$^{-1}$. For SC data, with a time sampling of about 1\,min, frequencies up to 720 \,d$^{-1}$ could be detected.

As a comparison, subsamples of the stars were analysed using different analysis methods, such as SigSpec (Reegen, 2007, 2011), Period04 (Lenz \& Breger 2005), the generalized Lomb-Scargle periodogram (Zechmeister \& K\"urster 2009), and 
the non-interactive code, \textit{\,freqfind\,} (Leroy \& Guti\'errez-Soto, in prep.). The latter code is based on the non-uniform fast Fourier transform by Keiner, Kunis \& Potts (2009), and significantly decreases the computation time for unevenly spaced data. The results obtained with the different methods were consistent. 

\section{Classification}
\subsection{\dsct, \gdor, and hybrid stars}
\label{sectclass}
We performed a careful inspection (one-by-one, and by eye) of the 750 light curves, the extracted frequency spectra, and list of detected frequencies, and tried to identify candidate \dsct, \gdor, and hybrid stars. We used a conservative approach and omitted frequencies with amplitudes lower than 20\,ppm for the classification. We also filtered out obvious combination frequencies and harmonics\footnote{As obvious combination frequencies and harmonics we considered $nf_i$ or $kf_i \pm lf_j$, with $f_i$ and $f_j$ different frequencies, $n \in [2, 3, 4, 5]$, and $k,l \in [1,2,3,4,5]$.} in an automatic way, and only considered apparent independent frequencies for the analysis. We suspect that the variable signal of a few stars is contaminated by the light variations of a brighter neighbouring star on the CCD. 
 We flagged all stars with a high contamination factor ($> 0.15$), as given by the KIC. If the light curves of the neighbouring stars on the CCD were available through KASOC\footnote{Unfortunately, only 40 stars of the sample could be checked in this way. We saw a clear contamination for the stars KIC\,4048488 and KIC\,4048494, KIC\,5724810 and KIC\,5724811, and KIC\,3457431 and KIC\,3457434. Less clear contamination is seen for KIC\,4937255 and KIC\,4937257, and KIC\,10035772 and KIC\,10035775, which are stars that show no obvious periodic variable signals.}, we carefully checked the light curves of these stars with their neighbours. Stars that show an obvious contamination effect were omitted from classification. We used information on  \Teff\,(Table\,\ref{tableTeff}) to distinguish between \dsct\,and \gdor\,stars versus $\beta$\,Cep and SPB stars. To be conservative, low frequencies ($<$0.5d$^{-1}$)  (see, for instance, the frequency spectra in Fig.\,\ref{lightcurvesGDOR}) are currently not taken into account in the analysis, because in this frequency range real stellar frequencies are contaminated with frequencies resulting from instrumental effects (see Sect.\,\ref{secttreatment}), and the separation of the different origins requires a dedicated study, which is beyond the scope of this paper.

We encountered a variety of light curve  behaviour. Based on a small number of stars and using only the first quarter of {\it Kepler} data, Grigahc\`ene et al. (2010) already proposed a subdivision of the AF-type pulsators into pure \dsct\,stars, pure \gdor\,stars, \dsct/\gdor\,hybrids and \gdor/\dsct\,hybrids, using the fact that frequencies are only detected in the \dsct\, (i.e.  $> 5$\,d$^{-1}$, or $> 58\,\mu$Hz) or \gdor\, (i.e.  $< 5$\,d$^{-1}$ or $< 58\,\mu$Hz) domain, or in both domains with dominant frequencies in either the \dsct\,star or \gdor\,star region, respectively. Among the 750 sample stars we see different manifestations of hybrid variability. There are stars that show frequencies with amplitudes of similar height in both regimes, and stars with dominant frequencies in the \gdor\,(\dsct) domain and low amplitude frequencies in the \dsct\,(\gdor) domain. The light curves show diversity as well. Balona et al. (2011d) already commented  on the different shapes of light curves of pure \gdor\,stars.

 In this work, we focus on stars that show at least three independent frequencies.  We classified the stars in three groups: \dsct\,stars,  \gdor\,stars, and hybrid stars. Because the underlying physics that causes the different types of hybrid behaviour is currently not clear, all types of hybridity (both \dsct/\gdor\,hybrids and \gdor/\dsct\,hybrids) are included in the group of hybrids. A star was classified as a hybrid star only if it satisfied all of the following criteria:
\begin{itemize}
\item Frequencies are detected in the \dsct\, (i.e.  $> 5$\,d$^{-1}$ or $> 58\,\mu$Hz) and \gdor\,domain (i.e.  $< 5$\,d$^{-1}$ or $< 58\,\mu$Hz),
\item The amplitudes in the two domains are either comparable, or the amplitudes do not differ more than a factor of 5--7 (case-to-case judgement), 
\item At least two independent frequencies are detected in both regimes with amplitudes higher than 100\,ppm.
\end{itemize}
By using these criteria, we should reduce the number of false positive detections. In particular, we tried to avoid a hybrid star classification of 'pure' \dsct\,stars that show a prominent long-term variability signal caused by rotation. We also tried to take care of more evolved \dsct\,stars that are expected to pulsate  with frequencies lower than 5\,d$^{-1}$. Stars that  exhibited only or mainly  frequencies in the \dsct\,domain (i.e.  $> 5$\,d$^{-1}$) and did not satisfy all of the above given criteria were assigned to the pure \dsct\,group. Likewise, the group of pure \gdor\,stars consists of stars that do not comply with the hybrid star criteria, and that have only or mainly frequencies lower than 5\,d$^{-1}$. However, the classification of pure \gdor\,stars is not as straightforward, because several other physical processes and phenomena can give rise to variability on similar timescales, such as binarity and rotational modulation caused by migrating star spots. We tried our best to select only \gdor\,stars, but are aware that nonetheless, and most likely, our selection is contaminated with a few non-{\it bona fide} \gdor\,stars. For stars that were observed in non-consecutive {\it Kepler} quarters, we tried to beware of frequencies introduced by the spectral window. For instance, frequently a peak near 48\,d$^{-1}$ (555 $\mu$Hz) is detected (e.g. KIC\,2166218 and KIC\,7798339), which for a \gdor\,pulsator can result in an incorrect classification as hybrid star. 

\begin{figure*}
\resizebox{0.99\linewidth}{!}{\rotatebox{-90}{\includegraphics{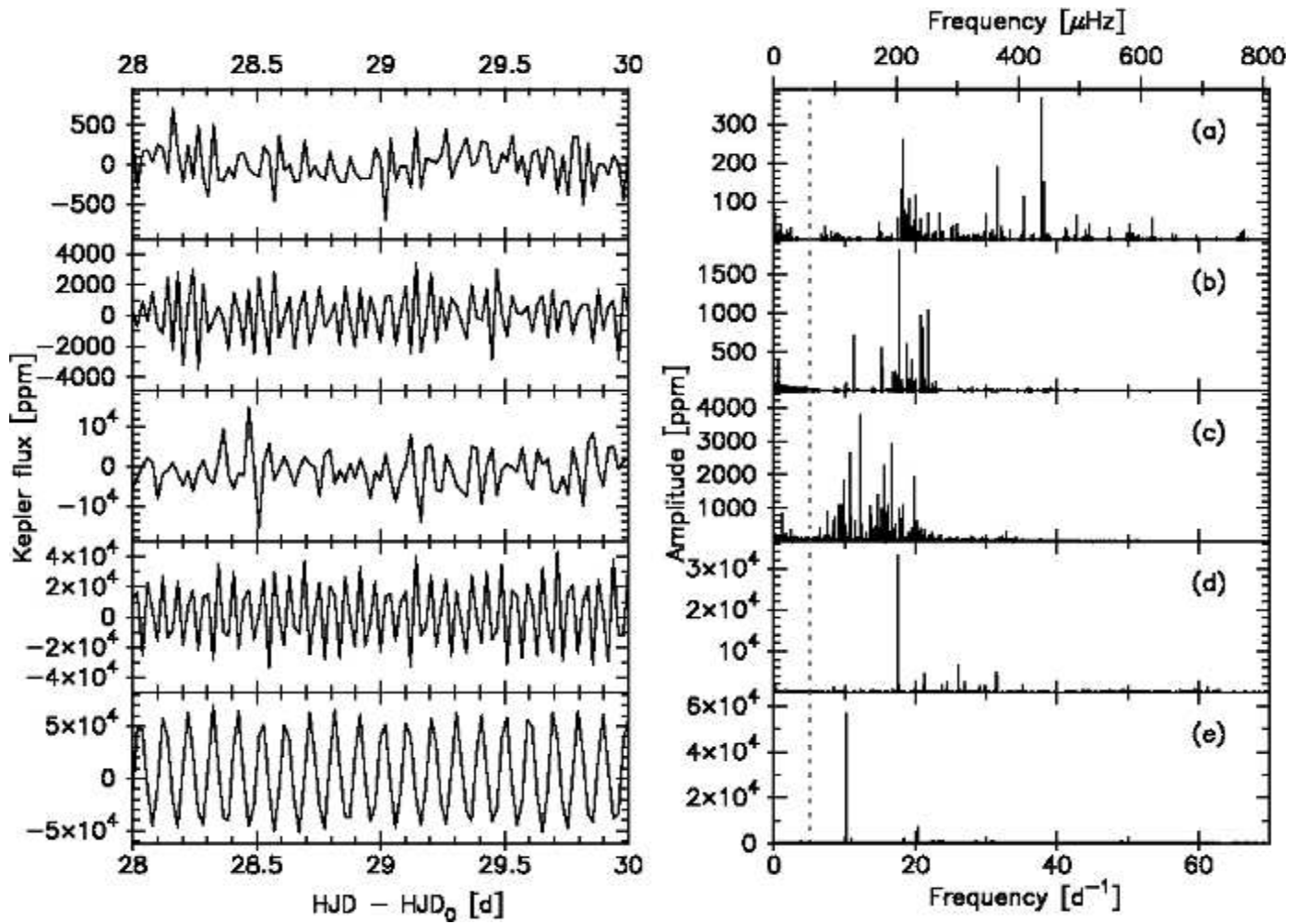}}}
\caption{Light curve and frequency spectrum of five stars assigned to the \dsct\,group, illustrating the variety of pulsational behaviour within the group. The left panel shows a portion of the {\it Kepler} light curves. The {\it Kepler} flux is expressed in ppm, HJD is given in d with respect to HJD$_0 = 2454950.0$. The right panel gives a schematic representation of the detected independent frequencies,  expressed in d$^{-1}$ (bottom X-axis) or $\mu$Hz (top X-axis). Amplitudes  are given in ppm. The dotted grey line separates the \dsct\,and \gdor\,regime. Note the different Y-axis scales for each star. (a) KIC\,8415752; (b) KIC\,8103917; (c) KIC10717871; (d) KIC\,9845907; (e) KIC\,9306095.}
\label{lightcurvesDSCT}
\end{figure*}

\begin{figure*}
\resizebox{0.99\linewidth}{!}{\rotatebox{-90}{\includegraphics{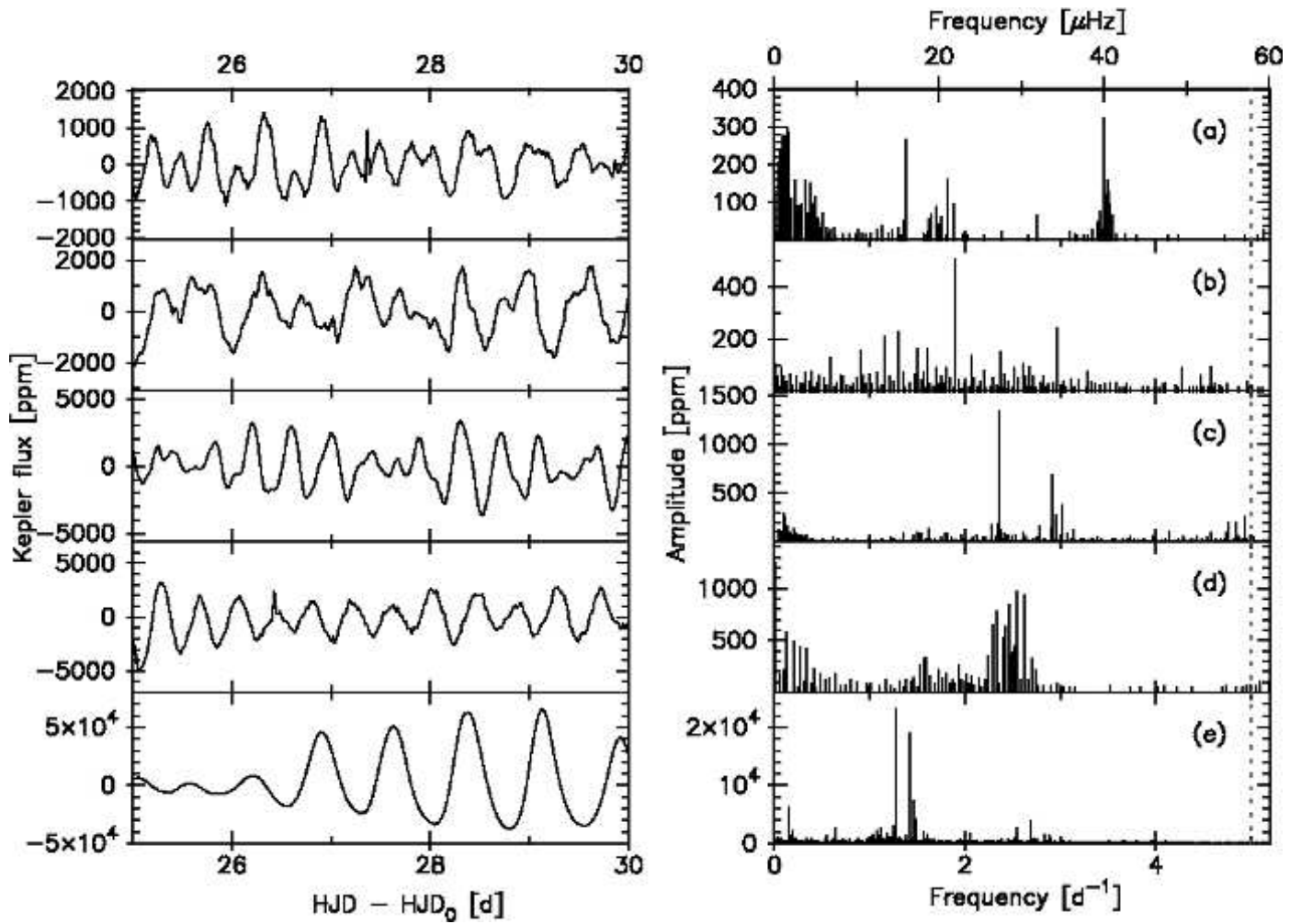}}}
\caption{Similar figure as Fig.\,\ref{lightcurvesDSCT}, but for five candidate \gdor\,stars. Note the different X-axis scale with respect to Fig.\ref{lightcurvesDSCT}. (a) KIC\,1432149; (b) KIC\,5180796; (c) KIC\,7106648; (d) KIC\,8330056; (e) KIC\,7304385.}
\label{lightcurvesGDOR}
\end{figure*}

\begin{figure*}
\resizebox{0.99\linewidth}{!}{\rotatebox{-90}{\includegraphics{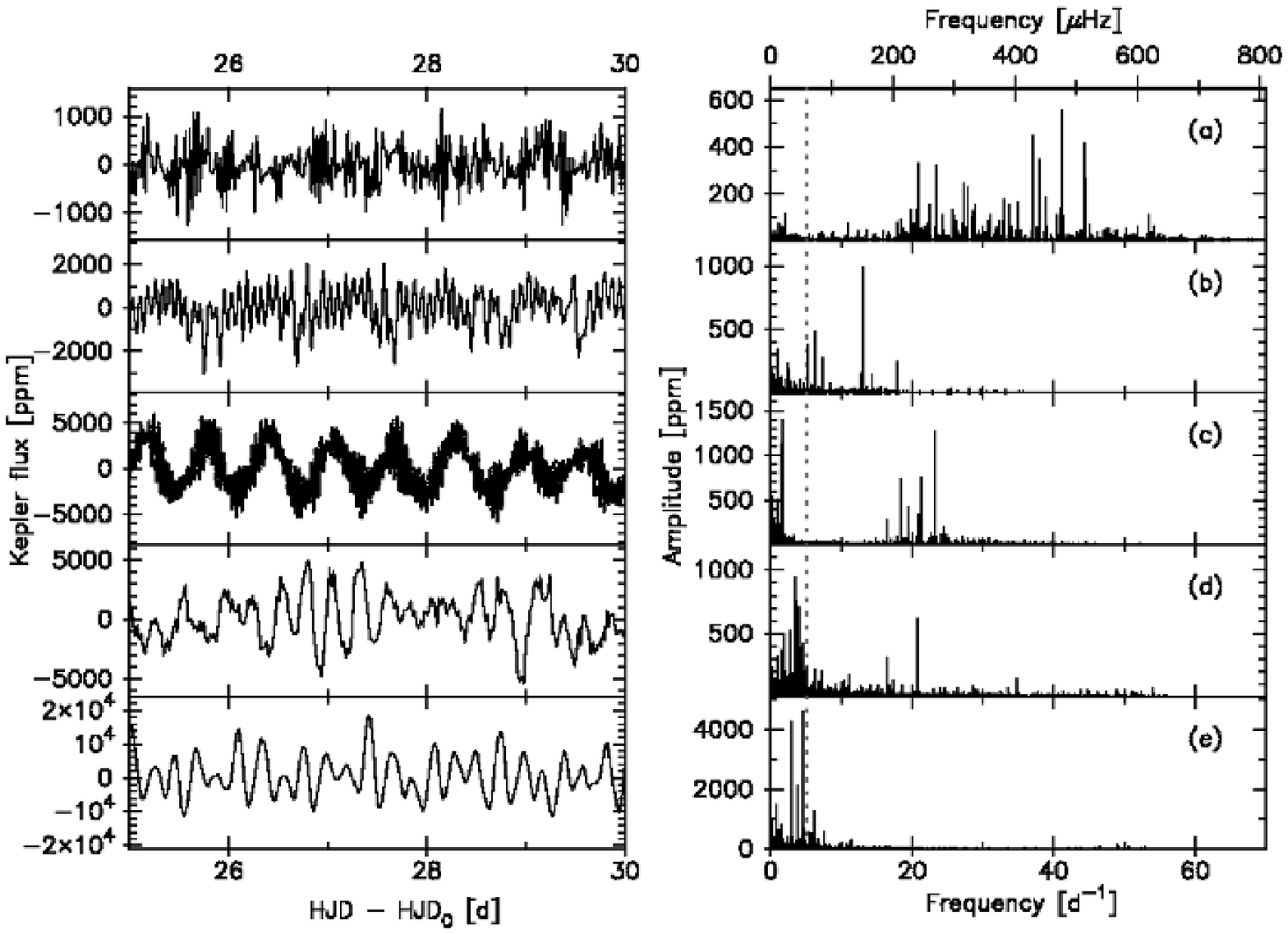}}}
\caption{Similar figure as Fig.\,\ref{lightcurvesDSCT}, but for five candidate hybrid stars. (a) KIC\,3119604; (b) KIC\,2853280; (c) KIC\,9664869; (d) KIC\,9970568; (e) KIC\,3337002. The two top panel (a)-(b) are \dsct\,frequency-dominated stars, and the two bottom panels (d)-(e) are \gdor\,frequency-dominated stars. }
\label{lightcurveshybrid}
\end{figure*}

In Figs\,\ref{lightcurvesDSCT}--\ref{lightcurveshybrid} a  portion of the light curve with a time span of 2\,d (\dsct\,stars) or 5\,d (\gdor\,and hybrid stars) and a schematic overview of the detected independent frequencies (i.e. combination frequencies are filtered out in an automated way, see above) are given for a few representative stars of each group. The amplitudes and {\it Kepler} flux are expressed in ppm, and the frequencies are given in both d$^{-1}$ (bottom X-axis) and $\mu$Hz (top X-axis). The dotted grey line in the amplitude spectra separates the \dsct\,and \gdor\,regime. The dates are in the Heliocentric Julian Date (HJD) format HJD$_0 = 2\,454\,950.0$. The figures illustrate the variety of pulsational behaviour within the groups.  
The \dsct\,stars (Fig.\,\ref{lightcurvesDSCT}) display an impressive variety of amplitude heights. The variability of the stars in panels (d) and (e),  KIC\,9845907 and KIC\,9306095, respectively, is dominated by one high-amplitude frequency. Several lower amplitude variations are also present. The chance of confusing a high amplitude \dsct\,star (HADS) and binarity is high for  KIC\,9306095. The stars in panels (a), (b), and (c) show multiperiodic variations with frequency amplitudes of similar size. The rotational frequency near 1.2 d$^{-1}$ and its first harmonic of the star KIC\,10717871  (panel c) could be mistaken for \gdor-like frequencies.  Because there are no other longer-term periodicities, there is no evidence for the possible hybrid status of this star (panel c).

The light curves of \gdor\,stars (Fig.\,\ref{lightcurvesGDOR}) vary from obvious beat patterns to less recognizable variable signals. Balona et al. (2011d) already pointed out that there are symmetric (e.g. panel d) and asymmetric (e.g. panel e) light curves among the stars that show obvious beating, and that most likely in these cases the pulsation frequencies are comparable to the rotation frequency.  Balona et al. (2011d) also suggested that the more irregular light curves likely stem from slowly rotating stars.  

Examples of hybrid stars are given in Fig.\,\ref{lightcurveshybrid}. The grey dotted line in the right panels guide the eye to separate the \dsct\,and \gdor\,regimes. The stars KIC\,3119604 and KIC\,2853280 (panels a and b, respectively) are clearly dominated by \dsct\,frequencies, while the \gdor\,frequencies have lower amplitudes. The star KIC\,9664869 (panel c) is an example of a star that exhibits frequencies with amplitudes of comparable height in the two regimes. The highest peak in the \gdor\,region is most likely related to the stellar rotation period, however, because several harmonics are also observed. The bottom two panels are examples of hybrid stars dominated by \gdor\,periodicities.

\addtocounter{table}{1}
Table\,\ref{classification},  available in the on-line version of the paper,  presents an overview of the stars  assigned to the three groups. For each star (KIC ID) we provide the classification (Class), the total number of independent frequencies (N) detected above the significance level (fap $= 0.001$) and with amplitudes higher than 20\,ppm, and the number of independent frequencies detected in the \gdor\,and \dsct\,regime (N$_{\rm \gamma Dor}$ and N$_{\rm \delta Sct}$, respectively). The next column gives as a reference the total number of frequencies detected above the significance level, including combination frequencies and harmonics (N$_{\rm total}$). The next four columns denote the frequency range of peaks in the \gdor\,and \dsct\,regimes ((Freq Range)$_{\rm \gamma Dor}$ and (Freq Range)$_{\rm \delta Sct}$, expressed in d$^{-1}$), the highest amplitude (Amplitude$_{\rm high}$, expressed in ppm) and associated frequency (Freq$_{\rm high}$, in d$^{-1}$). In the last column a flag ($\bullet$) indicates if the risk on light contamination with a neighbouring star on the CCD is high (contamination factor $> 0.15$). A typical error on the frequency associated with the highest amplitude is 0.0001\,d$^{-1}$. The error on the amplitude ranges from a few ppm up to about 30\,ppm. We note that for stars identified as  \gdor\,or \dsct\,stars we report on frequencies up to 6\,d$^{-1}$ or from 4\,d$^{-1}$, respectively, to account for, for instance, the frequency spectrum of more evolved stars. 

We note that for several stars classified as \gdor\,star only LC data are available. This may create a selection effect, because short-term \dsct\,periods are more difficult to detect in the  short timestring of LC data owing to sampling restrictions. Also, as mentioned above, even though we carefully checked the stars one by one, we expect to have a few false positive detections of hybrid and \gdor\,stars because the typical \gdor\,frequencies can be easily  confused with variations of the order of a day caused by rotation or  binarity.  A more careful analysis and interpretation of the full frequency spectrum of all individual stars of the sample will clarify this matter, but this is beyond the scope of this paper.

We compared our classification with the automated supervised classification results presented by Debosscher et al. (2011). Because these authors  studied public  {\it Kepler} Q1 data, only 479 objects of our sample appear in their catalogue.  We point out that the classifier by  Debosscher et al. (2009) only takes three independent frequencies with the highest amplitudes into account. Hence, the recognition and classification of hybrid behaviour is currently not implemented. Moreover, because the classifier does not take external information into account that can distinguish between B-type stars and AF-type stars (e.g. colour information, spectral classification based on spectra), there is often a confusion between \dsct\,and $\beta$\,Cep stars, and between \gdor\,and SPB stars. In general, there is good agreement ($> 87\%$, classified in terms of \dsct\,or $\beta$\,Cep stars) with the classification by  Debosscher et al. (2011) for stars that we classified as \dsct\,stars. The \gdor\,stars, as we classified them, are in general less easily recognized by the automated classifier. Often they appear as 'miscellaneous' in their list. This is not surprising, because so far only a few high-quality light curves of well-recognized \gdor\,stars were available that could be used as a template to feed the classifier. Stars that we identified as hybrid stars appear in the catalogue by Debosscher et al. (2011) as 'miscellaneous' or as \dsct, \gdor, $\beta$\,Cep, or SPB stars. The work presented in this paper will provide valuable feedback and information to refine the automated supervised classification procedure developed by Debosscher et al. (2009).


\subsection{Other  classes}
\label{sectotherclass}
\begin{figure*}
\resizebox{0.99\linewidth}{!}{\rotatebox{-90}{\includegraphics{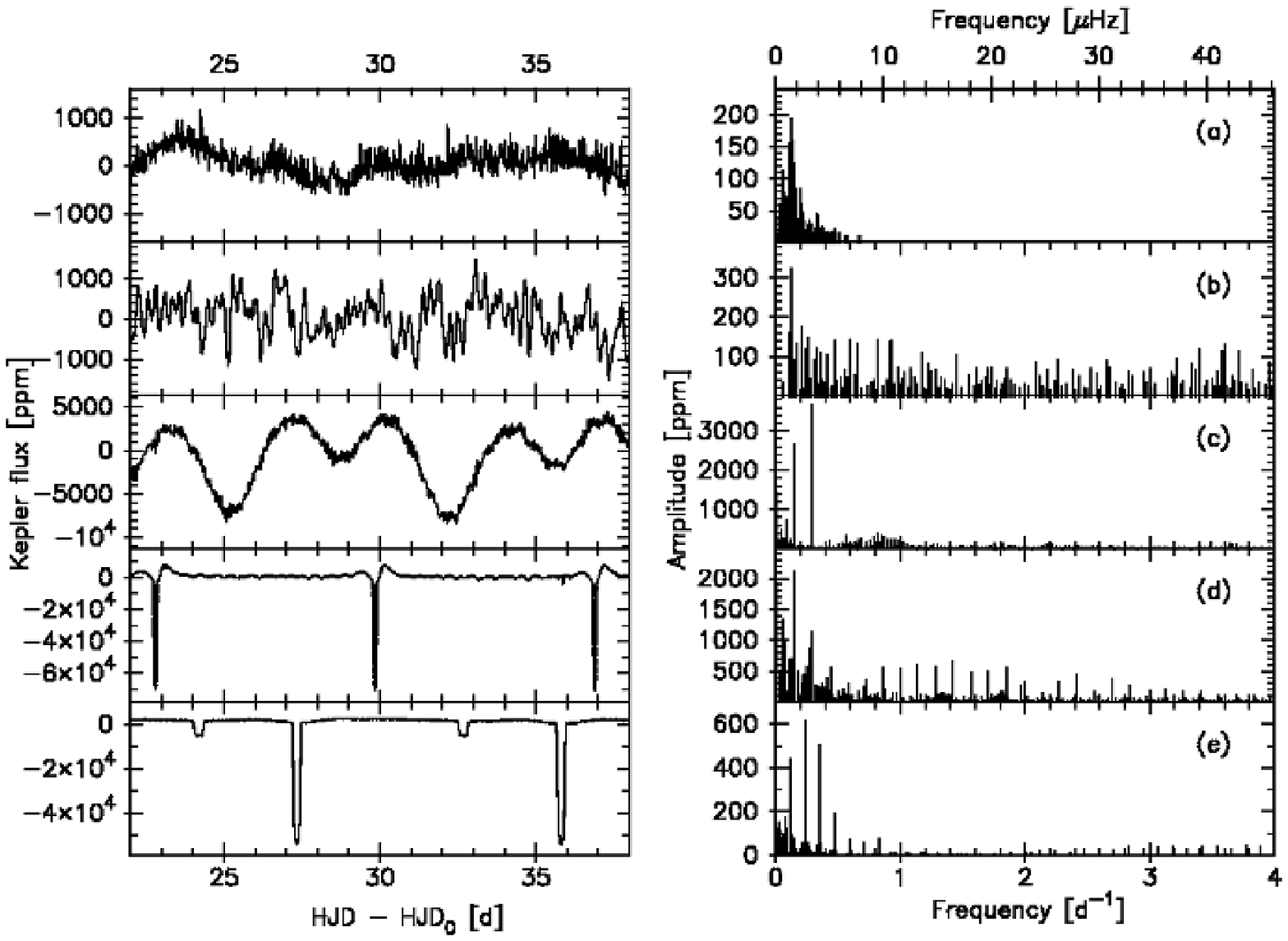}}}
\caption{Similar figure as Fig.\,\ref{lightcurvesDSCT}, but for stars that were not assigned to the groups of \dsct, \gdor\,or hybrid stars. (a) KIC\,9386259, no clear periodic signal detected; (b) KIC\,2584202, red giant star;
 (c) KIC\,5197256, EB or ellipsoidal variable with a \dsct\,component; (d) KIC\,3230227, EB with a \gdor\,component; (e) KIC\,9851142, EB with most likely a \gdor\,component.}
\label{lightcurvesNA+bin}
\end{figure*}

\begin{figure*}
\resizebox{0.99\linewidth}{!}{\rotatebox{-90}{\includegraphics{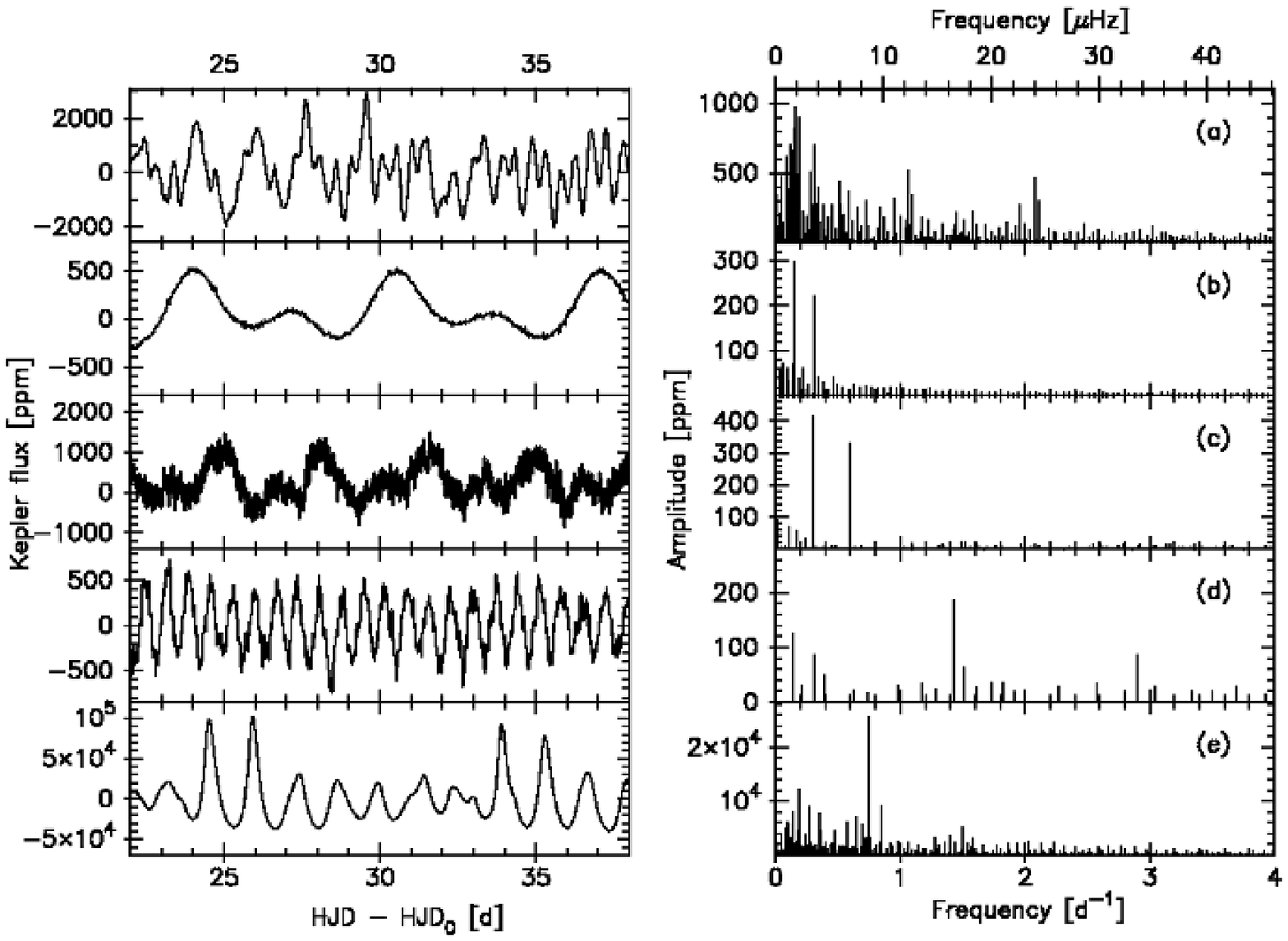}}}
\caption{Similar figure as Fig.\,\ref{lightcurvesDSCT}, but for stars that were not assigned to the groups of \dsct, \gdor\,or hybrid stars. Stellar activity/rotational modulation: (a) KIC\,8748251, (b) KIC\,8703413 and (c) KIC\,11498538; No clear classification: (d) KIC\,12062443, and (e) KIC\,3348390.}
\label{lightcurvesother}
\end{figure*}

\addtocounter{table}{1}
About 63\% of our sample is recognized as \dsct, \gdor, or hybrid star. 
Table\,\ref{other}, in the on-line version of the paper,  gives an overview of the 'classification' of the remaining 37\% of the stars.  For each star (KIC ID) the associated classification (Class) and a flag (Flag) indicating a high risk on light contamination by a neighbouring star ($\bullet$ if there is a contamination factor $> 0.15$), are given. Table\,\ref{other}  includes stars that show no clear periodic variability on timescales typical for \dsct\,and \gdor\,pulsators ('\ldots', or 'solar-like'), stars that exhibit stellar activity and show a rotationally modulated signal ('rotation/activity'), binaries ('binary' or  eclipsing binary 'EB'), B-type stars ('Bstar'), candidate red giant stars ('red giant'),  Cepheids ('Cepheid'), and stars whose light is contaminated by another star ('contaminated'). Although the observed ranges in \Teff\,and \logg\, include typical values for RR\,Lyr stars (see Fig.\,\ref{750stars}), we did not find any in our sample, but there are $\sim$40 such stars observed by {\it Kepler}, which are studied separately (Kolenberg et al 2010; Benk{\H o} et al. 2010). Unclear cases mostly show a behaviour that might be related to rotation and are hence also labelled 'rotation/activity'. We also assigned the candidate \gdor\,stars for which less than three significant peaks were detected to this category. The light curve and frequency spectra of a few examples of these other classifications are given in Figs\,\ref{lightcurvesNA+bin} and \ref{lightcurvesother}.

One hundred and twenty-one stars do not show an obvious periodicity in the expected range for \gdor\,and \dsct\,stars, or have an unresolved frequency spectrum within the available dataset.  The star KIC\,9386259 (Fig.\,\ref{lightcurvesNA+bin}, panel a) is an example of a star showing no clear periodicity.  Furthermore,  we used the label '\ldots' for some stars for which less than three significant frequencies were detected (e.g. KIC\,11509728 and KIC\,11910256). We investigated the stars for signatures of solar-like oscillations and identified 75 candidate solar-like oscillators ('solar-like', see Table\,\ref{other}). 

We identified seven B-type stars and 44 red giant stars in the sample. The giant stars show an envelope of frequencies with amplitudes up to 100--200\,ppm in the region 0.5-5\,d$^{-1}$, as illustrated by KIC\,2584202 (Fig.\,\ref{lightcurvesNA+bin}, panel b).  Among the B-type stars, we recognized five SPB stars and one candidate \bcep\,star. 

Within the sample we identified at least 39 binaries, including 28 EBs. In Table\,\ref{other} the binary stars are labelled 'binary', or 'EB'  for an EB. If the variability of one of the components is identified as typical for one of the three groups outlined in Sect.\,\ref{sectclass}, we also indicated this in Table\,\ref{other}. Panels (c), (d) and (e) of Fig.\,\ref{lightcurvesNA+bin} show examples of EBs. An interesting target is KIC\,11973705, because it most likely is a binary with a \dsct\,and SPB component (see also Balona et al. 2011b). For three stars reported in the literature as EBs (Pr\v{s}a et al. 2011; Slawson et al. 2011; Hartman et al. 2004), KIC\,2557115, KIC\,5810113, and KIC\,1432149,  we find no clear evidence of their eclipsing nature in the {\it Kepler} lightcurves.  In case of KIC\,1432149, presented by Hartman et al. (2004) as an EB with period 9.3562\,d, we cannot confirm its eclipsing nature or its orbital period, and we suspect that this target has been misidentified as an EB.

Several stars show an irregular light curve typical of stellar activity, or a clearly rotationally modified signal (panels (a)-(c) of Fig.\,\ref{lightcurvesother}). It is also not impossible that low-amplitude pulsating \gdor\,star candidates are hidden among the stars labelled as 'rotation/activity' in Table\,\ref{other}. Namely, when only one or two of their pulsation frequencies reach the current detection threshold, they are not yet assigned to a pulsation group. A possible \gdor\,candidate is given in panel (d) of Fig.\,\ref{lightcurvesother}.

In some cases the light curves look very peculiar, and the origin of the variability is not clear. This is the case for KIC\,3348390 (panel (e) of Fig.\,\ref{lightcurvesother}) and KIC\,4857678, for instance.

We discovered several interesting targets among the 750 stars of the sample. Dedicated studies of groups of individual stars will be presented in forthcoming papers. Below, we will sort the stars into different classes.

\section{Characterization of the different classes}
\label{Sectcharacterization}
The classification described in the previous section results in the following distribution. A total of 63\% of the sample can be identified as \gdor, \dsct\,or hybrid stars: 27\% are classified as \dsct\,stars (206 stars), 23\% as hybrid stars (171 stars; of which 115 stars are \dsct-dominated and 56 stars are \gdor-dominated), and 13\% as \gdor\,stars (100 stars).  A striking result is that almost a quarter of the sample, i.e. 171 stars, shows hybrid behaviour. This is in sharp contrast with the results obtained from ground-based observations, where so far only three candidate \gdor-\dsct\,hybrid stars have been discovered. The far superior precision of the space data opens a new window in detecting low amplitude variations. This result was already hinted at by Grigahc\`ene et al. (2010) and Hareter et al. (2010), but the quantification by means of this sample is remarkable. 

Of the remaining 37\% of the sample, a considerable number (121 stars, 16\%) do not show clear variability with periods in the expected range for \gdor\,and \dsct\,stars. Among this group are 75 candidate solar-like oscillators. Our sample has seven B-type stars (1\%) and 44 stars (6\%) are identified as red giant stars. One Cepheid turned out to be among the sample. About 8\% of the sample shows stellar activity, often manifesting itself by a rotationally modulated signal.

At least 5\% of the sample stars are identified through the analysis of their light curve as binary or multiple systems, of which 3.5\% show eclipses. When we also consider the known binaries from the literature  (Table\,\ref{database}), we arrive at a binary rate of 12\% within the sample. The number of binary detections is only a fraction of what is expected. The binary rate among A-F type stars in general and \dsct\,stars in particular is estimated to be at least 30\% (Breger \& Rodr\'{\i}guez 2000; Lampens \& Boffin 2000).  Several additional stars are expected to be part of multiple systems with possibly much longer periods than the available {\it Kepler} time span.  The percentage of EBs in our sample is high. Pr\v{s}a et al. (2011) reported a 1.2\% occurence rate of EBs among the {\it Kepler} targets.

\begin{figure}
\centering
\resizebox{0.90\linewidth}{!}{\rotatebox{-90}{\includegraphics{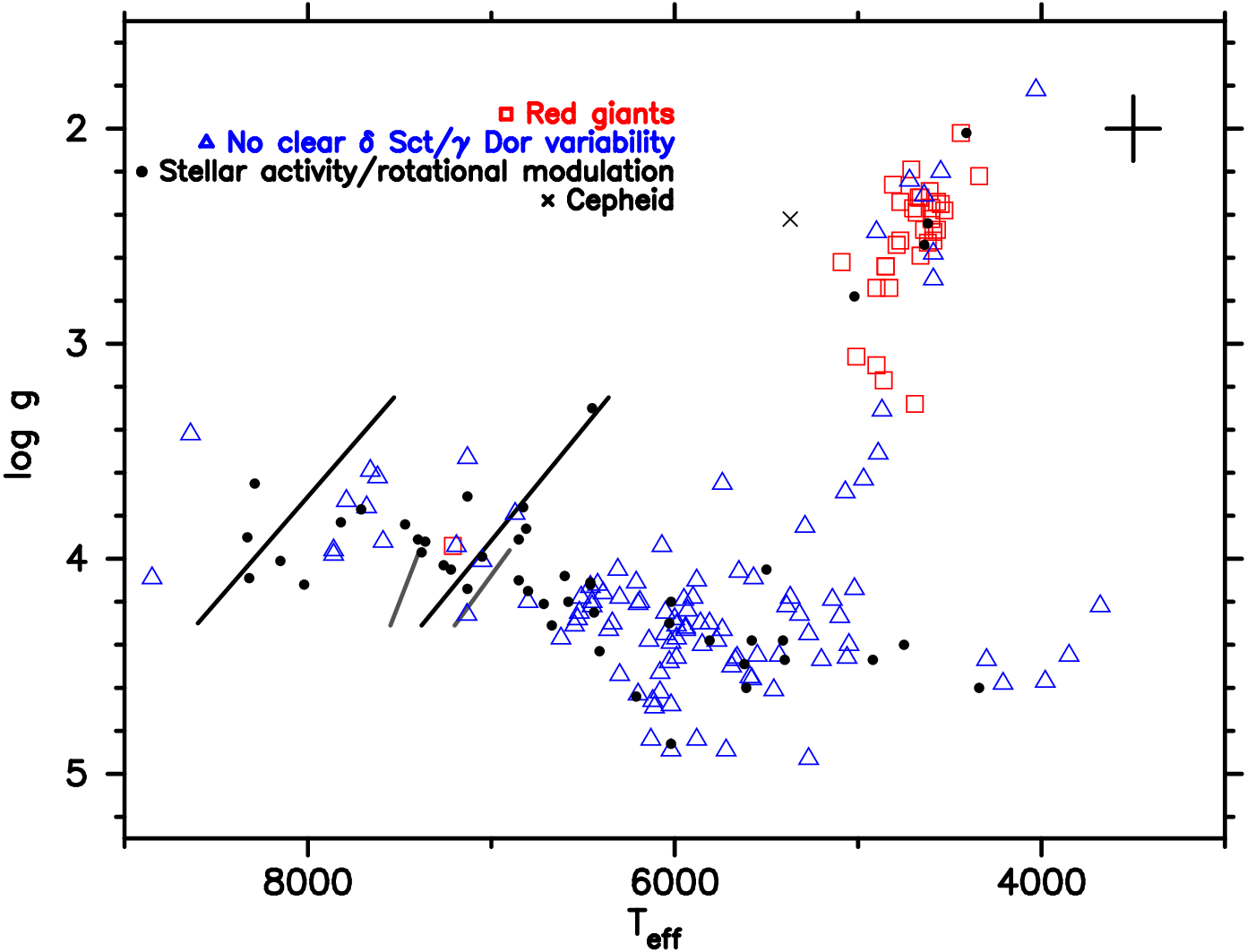}}} 
\caption{(\Teff, \logg)-diagram with stars that show no clear periodic variability on timescales typical for \dsct\,and \gdor\,pulsators (open triangles), stars identified as red giants (open squares), stars that exhibit stellar activity (bullet), and a Cepheid (cross).  The cross at the right top corner represents the typical error bars on the values:  290\,K for \Teff\,and 0.3\,dex for  \logg. The solid thick black and light grey  lines mark the blue and red edge of the observed instability strips of \dsct\,and \gdor\,stars, as described by Rodr\'{\i}guez \& Breger (2001) and Handler \& Shobbrook (2002), respectively. In the on-line version of the paper the open squares, open triangles, bullets, and crosses, are red,  blue, black, and  black, respectively.}
\label{Teffloggother}
\end{figure}

Figure\,\ref{Teffloggother} shows the stars that are not assigned to one of the \dsct, \gdor\,or hybrid star groups in a (\Teff, \logg)-diagram. The solid thick black and light grey lines mark the blue and red edge of the observed instability strip of \dsct\,and \gdor\,stars, respectively (Rodr\'{\i}guez \& Breger 2001; Handler \& Shobbrook 2002). Owing to the possibly incorrect separation of the binary component's contribution, we considered the physical parameters of the binaries as insufficiently constrained and omitted them. The same holds for the B-type stars, which are much hotter than the \Teff\,region shown here. The stars that show no clear periodic variability on timescales typical for \dsct\,and \gdor\,pulsators (open triangles) and stars that exhibit stellar activity (bullets) are found along the MS and in more evolved stars. The location of the only Cepheid in our sample is marked by a cross. The candidate red giants (open squares) are all but one found in the expected region of the (\Teff, \logg)-diagram. This implies that the KIC photometry separates giant from MS stars well.

\subsection{Characterization of stars that show no clear periodic variability}
\begin{figure}
\centering
\begin{tabular}{cc}\setlength{\tabcolsep}{0.1pt}
\hspace{-4mm}\resizebox{0.49\linewidth}{!}{\includegraphics{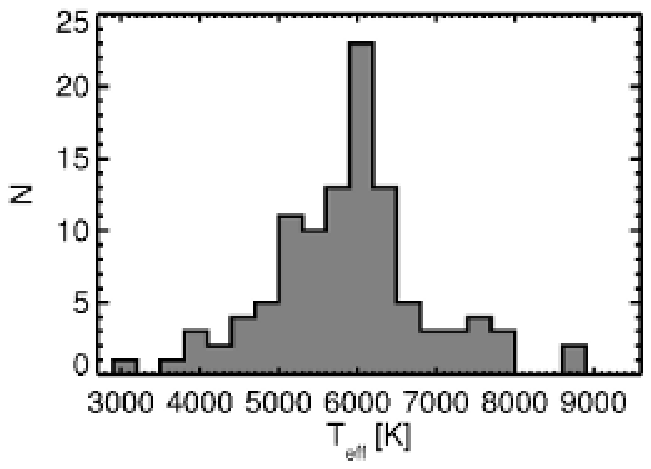}}&
\hspace{-4mm}\resizebox{0.49\linewidth}{!}{\includegraphics{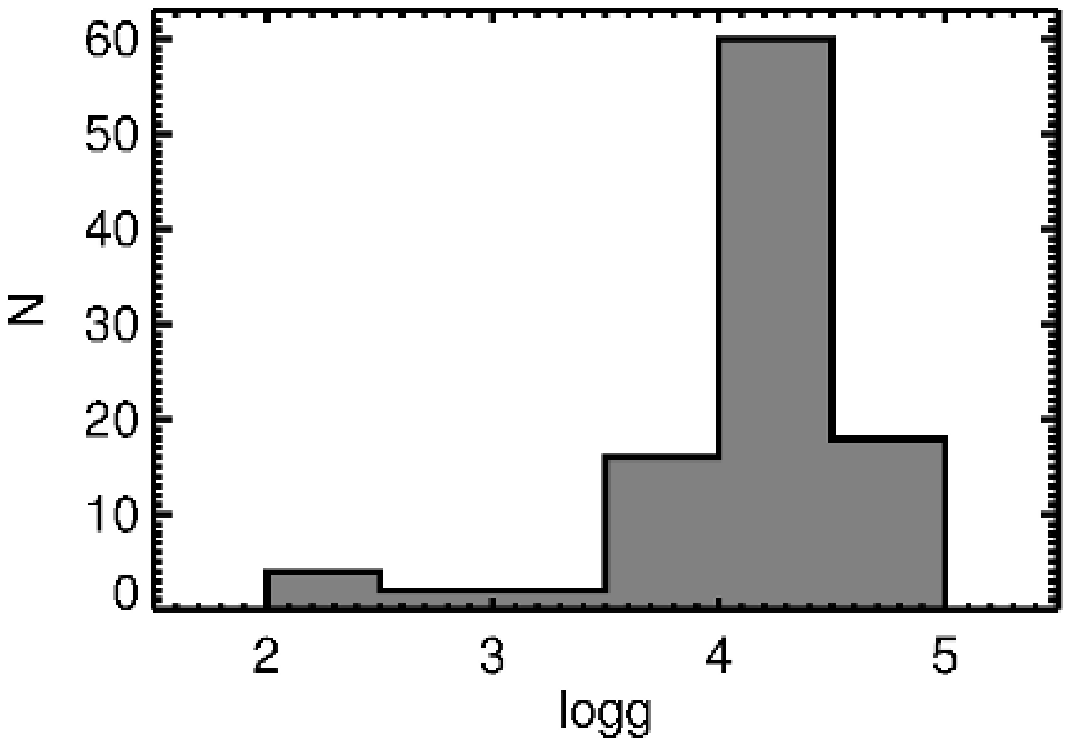}}
\\
\hspace{-4mm}\resizebox{0.49\linewidth}{!}{\includegraphics{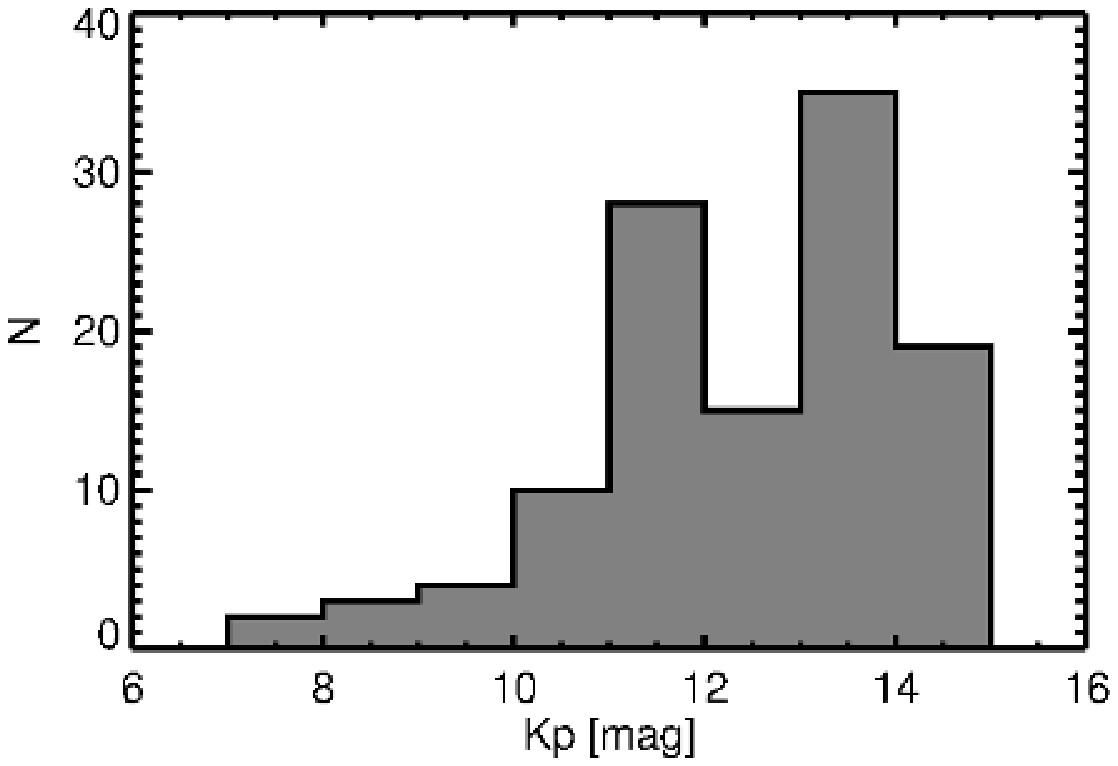}}&
\hspace{-4mm}\resizebox{0.49\linewidth}{!}{\includegraphics{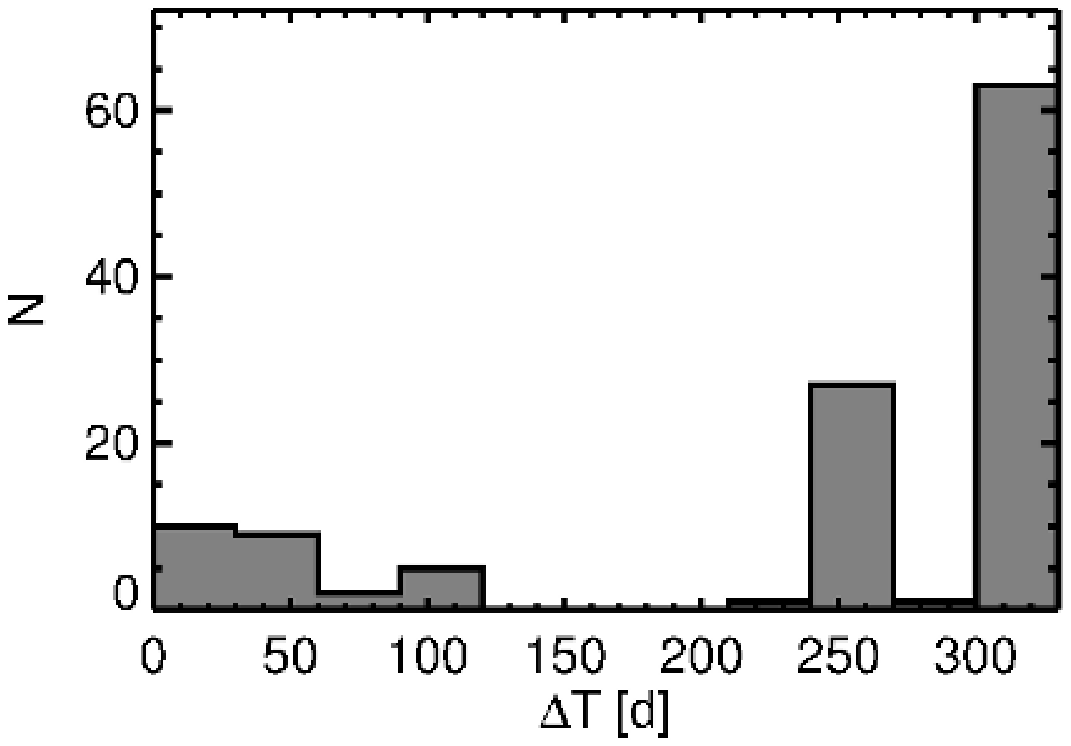}}\\
\end{tabular}
\caption{Distribution in \Teff\,(top left), \logg\,(top right), {\it Kepler} magnitude Kp (bottom left), and total time span $\Delta$T of the {\it Kepler} light curves  (bottom right)  of the 121 stars that show no clear periodic varibility. The number of stars belonging to each bin is given on the Y-axis.}
\label{NA}
\end{figure}

We now focus on the properties of the 121 stars that show no clear periodic variability in the \gdor\,and \dsct\,range of frequencies to understand why no oscillations are detected. Figure\,\ref{NA} presents the distribution in \Teff\,(top left), \logg\,(top right), Kp (bottom left), and total time span $\Delta$T, expressed in d, of the {\it Kepler} light curves (bottom right). 

The cool boundary of the observational instability strip for \gdor\,stars is located around T$_{\rm eff} = 6900$\,K. At least \footnote{For 11\% of the 121 stars we have no information on \Teff\,or \logg.} 78\% of the 121 stars have cooler temperatures, and hence  no A-F type variability is expected. About 75 stars are identified as candidate solar-like oscillators. However, 10\% of the 121 stars that show no clear periodicity are located inside the instability strip of \gdor\,or \dsct\,stars\footnote{As demonstrated in Sect.\,\ref{sectChar}, a revision of the current instability strip is required.} (see also Fig.\,\ref{Teffloggother}). Additional investigation is needed to confirm that these stars do not show variability, which would imply that non-variable stars exist in the instability strip. 

 Sixty percent (71 stars) of our non-variable stars are fainter than Kp\,$= 12$\,mag, and 18\% are fainter than  Kp\,$= 14$\,mag.  The faintness of the star most likely has an impact on the (non-)detection of periodicities. To quantify this, we  counted the fraction of apparently non-periodic stars per magnitude bin for the full sample of 750 stars. The number of stars that show no clear peridodicity increases dramatically towards faint stars: the fraction is only 2\% for magnitude Kp\,$= 9$\,mag, 5\% for Kp\,$= 10$\,mag, 12\% for Kp\,$= 11$\,mag, 15\% for Kp\,$= 12$\,mag, 41\% for Kp\,$= 13$\,mag, and 68\% for Kp\,$> 14$\,mag. The fainter the star, the more difficult it becomes to detect periodicities. Our analysis results, which were obtained by only considering amplitudes above 20\,ppm, lead us to  suspect that the {\it Kepler}  detection limit of A-F type low-amplitude oscillations ($\leq 20$\,ppm) lies around  Kp\,$= 14$\,mag (see also Sect.\,\ref{sectChar}). 

We find no evidence for a selection effect towards stars with a short time span in the available {\it Kepler} time series. The right panel of Fig.\,\ref{NA} shows that also several time series with long time spans do not show clear variability. Also, the observing mode has no obvious influence on the (non-)\,detection of oscillations. Fifty-four percent of the 121 stars have only LC data, while 46\% have only SC data.

To summarize, stars that show no clear periodic variations are generally the cooler and fainter stars of the sample. We do not find evidence for a bias towards the total time span of the available light curve or towards the observing mode (LC versus SC). 

\subsection{Characterization of \dsct, \gdor, and hybrid stars}
\label{sectChar}
\subsubsection{The (\Teff, \logg)-diagram}
\begin{figure*}
\centering
\begin{tabular}{cc}
\resizebox{0.44\linewidth}{!}{\rotatebox{-90}{\includegraphics{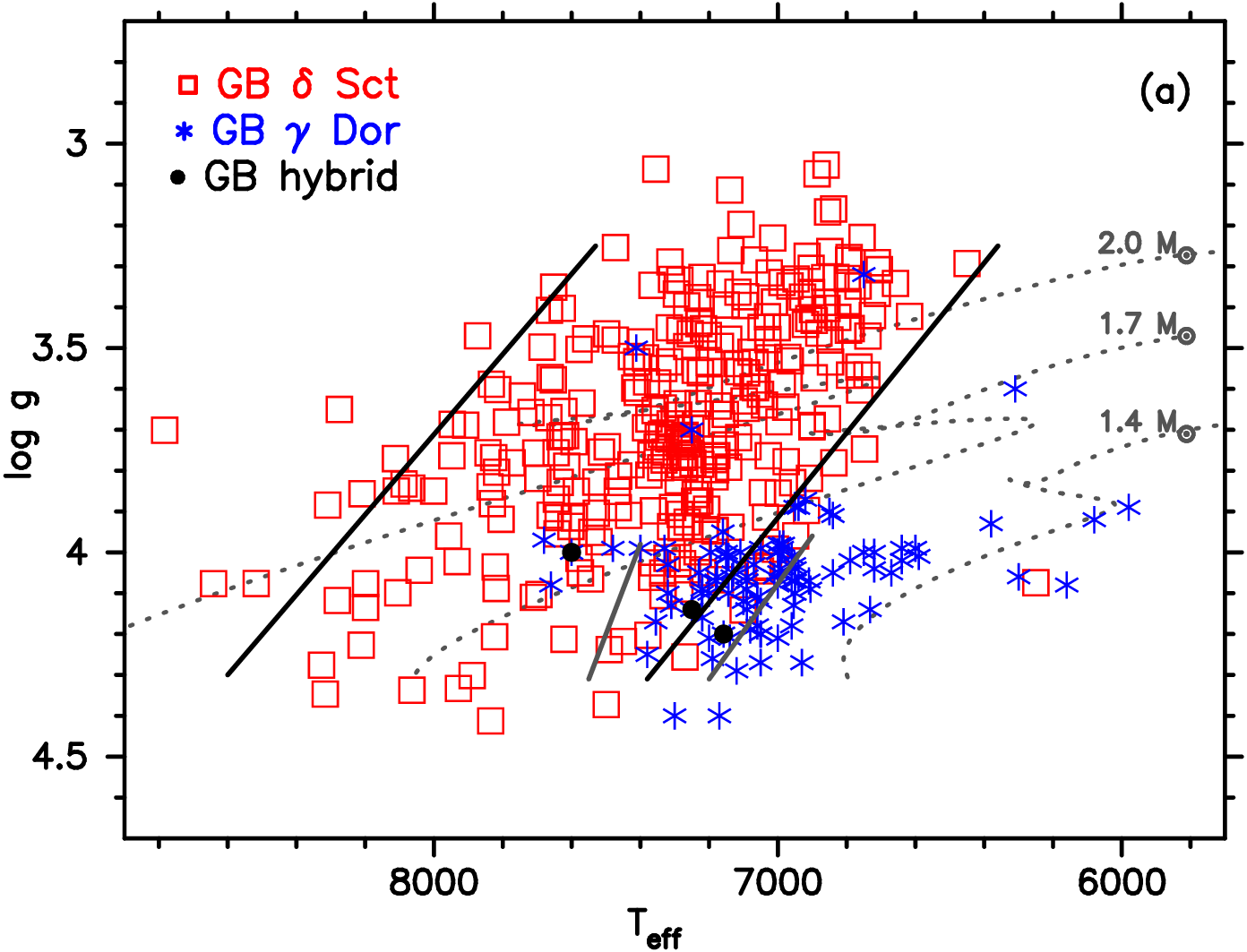}}} &
\resizebox{0.44\linewidth}{!}{\rotatebox{-90}{\includegraphics{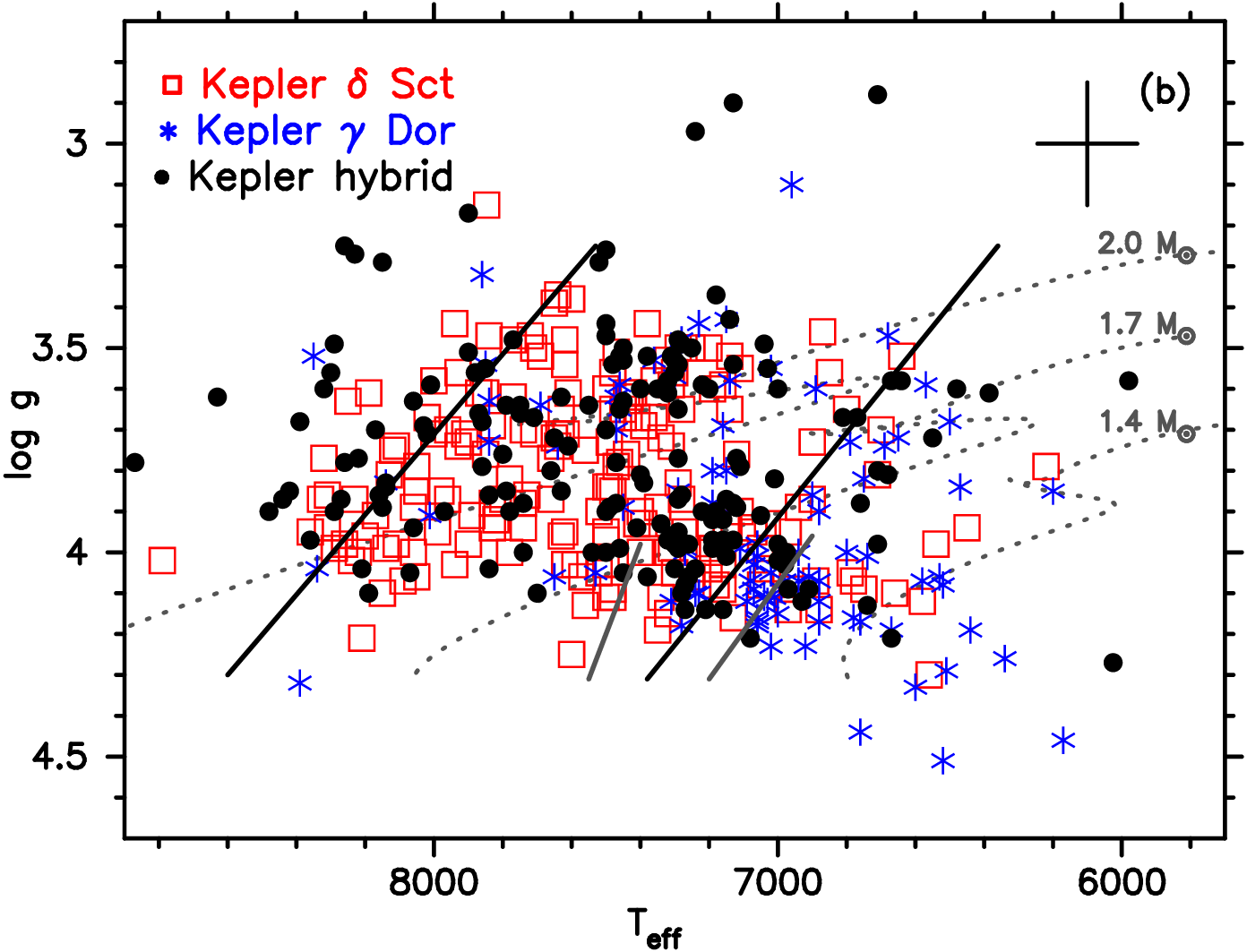}}} \\
\resizebox{0.44\linewidth}{!}{\rotatebox{-90}{\includegraphics{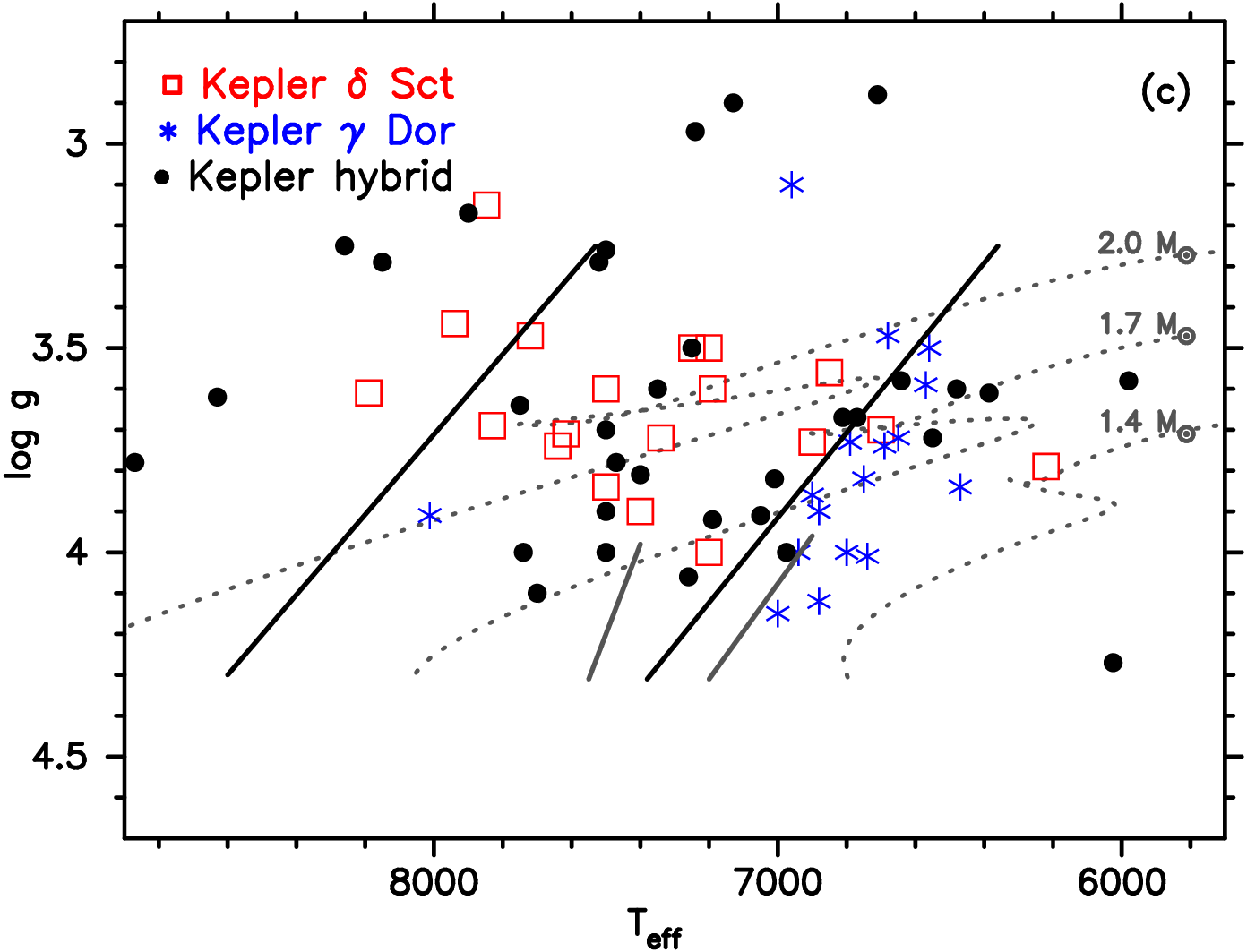}}} &
\resizebox{0.44\linewidth}{!}{\rotatebox{-90}{\includegraphics{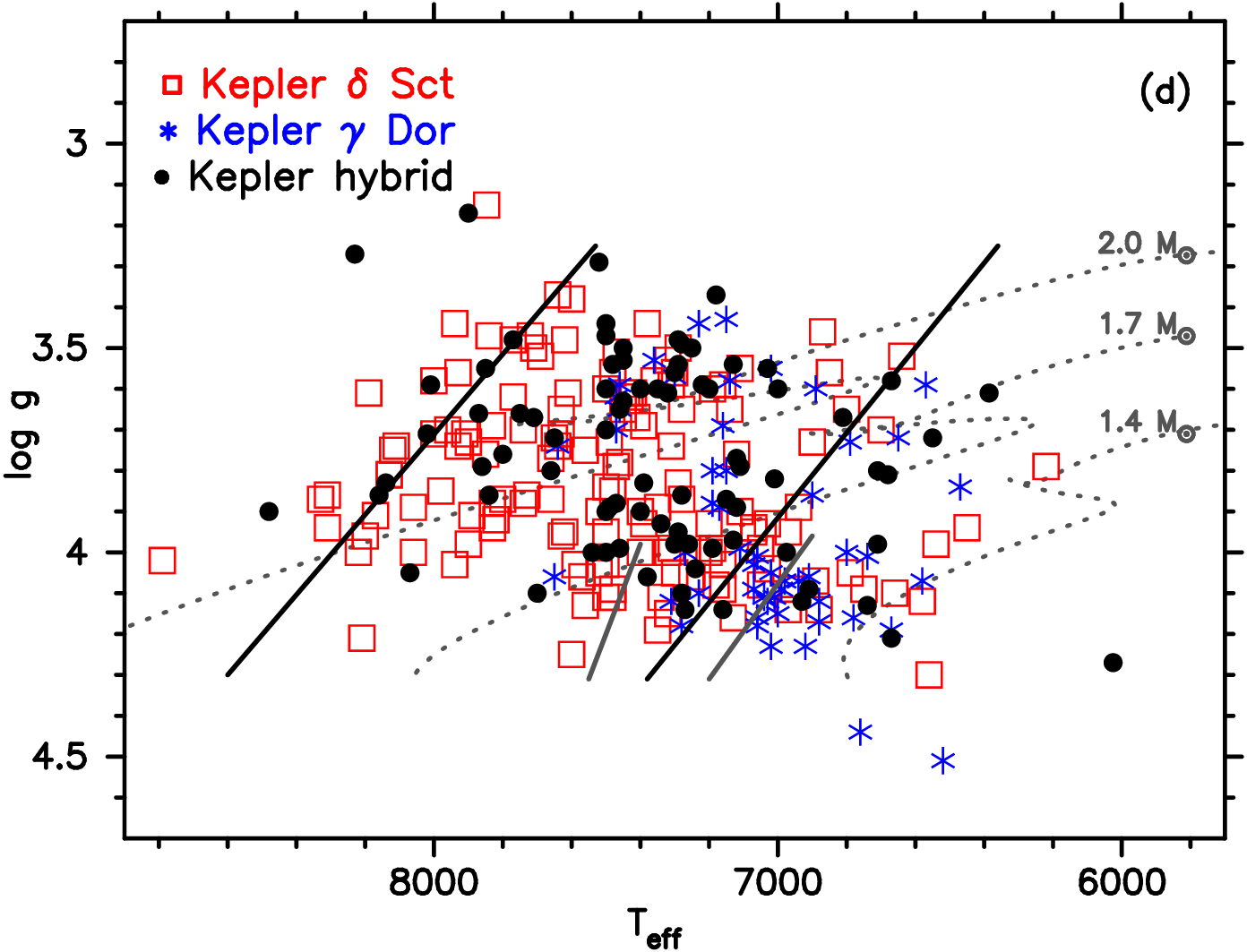}}}\\
\end{tabular}
\caption{(a): (\Teff, \logg)-diagram of the \dsct, \gdor, and hybrid stars detected from the ground (parameters taken from the literature). (b): (\Teff, \logg)-diagram of the {\it Kepler} stars we classified as  \dsct, \gdor, and hybrid stars in this paper. Open squares represent \dsct\,stars, asterisks indicate \gdor\,stars, and hybrid stars are marked by bullets. The black cross in the right top corner shows typical errors on the values. (c): (\Teff, \logg)-diagram of the subsample of 69 {\it Kepler} stars for which accurate \Teff\,and \logg\,values are available. The colour-codes are the same as for panel (b). (d) (\Teff, \logg)-diagram of the subsample of {\it Kepler} stars that show pulsations with amplitudes higher than 1000\,ppm ($> 1$\,mmag). Evolutionary tracks for MS stars with masses 1.4\,M$_{\odot}$, 1.7\,M$_{\odot}$, and 2.0\,M$_{\odot}$ are plotted with grey dotted lines. The evolutionary tracks  have been computed using the Code Li\'egeois d'\'Evolution Stellaire (CLES, Scuflaire et al. 2008). The input physics included is similar to the one used in Dupret et al. (2005) with the following values for the modelling parameters $\alpha_{\rm MLT} = 1.8$, $\alpha_{\rm ov} = 0.2$ and  $Z = 0.02$. The solid thick black and light grey  lines mark the blue and red edge of the observed instability strips of \dsct\,and \gdor\,stars, as described by Rodr\'{\i}guez \& Breger (2001) and Handler \& Shobbrook (2002), respectively. In the on-line version of the paper the symbols representing the \dsct, \gdor, and hybrid stars are red,  blue, and  black, respectively.}
\label{logTefflogg}
\end{figure*}

The current ground-based (GB) view on the positions of the \dsct\,and \gdor\,classes in the (\Teff, \logg)-diagram (parameters are taken from the literature\footnote{Rodr\'{\i}guez \& Breger 2001; Rodr\'{\i}guez et al. 2000; Henry \& Fekel (2005); Poretti et al. (1997); Breger et al. (1997); Zerbi et al. (1997, 1999); Aerts et al. (1998); Kaye et al. (1999b); Gray \& Kaye (1999); Eyer \& Aerts (2000); Guinan et al. (2001); Aerts (2001); Mart\'{\i}n et al. (2003); Mathias et al. (2004); Rowe et al. (2006);  Bruntt et al. (2008); Cuypers et al. (2009); Uytterhoeven et al. (2008); Catanzaro et al. (2010, 2011)}) is presented in panel (a) of Fig.\,\ref{logTefflogg}.  A comparison of \logg\,values derived from Geneva photometry and from other sources (photometry and spectroscopy) indicates a systematic difference of about 0.4\,dex for \logg\,values above 4.35\,dex as calculated from the Geneva photometry (Cuypers \& Hendrix, private communication). Therefore we have corrected the values based on Geneva photometry.  Evidently, the \dsct\,and \gdor\,stars occupy distinct locations in the (\Teff, \logg)-diagram, with a small overlap region. 

Panel (b) of Fig.\,\ref{logTefflogg} shows a different picture.
Here the \dsct, \gdor, and hybrid stars from the {\it Kepler} sample are plotted. We used the adopted values of \Teff\,and \logg, as given in Table\,\ref{tableTeff}. The cross in the top right corner of the figure shows typical errors on the values. The stars are scattered in the (\Teff, \logg)-diagram: the \dsct\,and \gdor\,stars are not confined anymore to the two regions that were clearly seen for the ground-based stars. Even when considering the large error bars on the values, the scatter is present. {\it Kepler} \dsct\,stars exist beyond the red edge of the instability strip, while {\it Kepler} \gdor\,pulsations appear in both hotter and cooler stars than previously observed from the ground. The {\it Kepler} hybrid stars occupy the entire region between the blue edge of the \dsct\,instability strip and the red edge of the \gdor\,instability strip, and beyond. The position of the {\it Kepler} \dsct\,and \gdor\,stars suggests that the edges of the so far accepted observational instability strips need to be revised. However, we need accurate values of \Teff\,and \logg\,for all stars to confirm this finding.

Because for most stars in our sample only KIC-based estimates of \Teff\,and \logg\,are available, we selected the stars that have reliable estimates of these parameters derived from ground-based spectra or multi-colour photometry (see Sect.\,\ref{sectphysparam}). From this selection, 69 are classified as belonging to one of the three groups. The subsample of 69 stars is plotted in panel (c) of  Fig.\,\ref{logTefflogg}. The position of the stars in the (\Teff, \logg)-diagram confirms the general findings described for the full sample. However, the scatter across the diagram of \gdor\,stars is less present, but almost all \gdor\,candidates lie outside the observational instability strip for \gdor\,stars.  Ground-based observations for the derivation of more precise values of \Teff\,and \logg\,are needed for all other stars to confirm the exact locations of the stars.

Panel (d) of Fig.\,\ref{logTefflogg} shows the {\it Kepler} stars assigned to the three groups that have amplitudes higher than 1000\,ppm (see Table\,\ref{classification}), which approximately corresponds to amplitudes higher than 1\,mmag and hence might be observable from the ground. We notice that the {\it Kepler} stars with ground-based observable amplitudes also do not fit within the observational instability strips.

The left column of Fig.\,\ref{histogram1} presents an overview of the distribution in \Teff\,for the three groups of A-F type stars. The histograms related to \dsct, hybrid, and \gdor\,stars are coloured in dark grey, middle grey, and light grey respectively. The distribution in \Teff\,peaks around 7400\,K, 7200\,K, and 7000\,K for \dsct, hybrid, and \gdor\,pulsators, respectively. Comparing these values with the center of the observed instability strips by Rodr\'{\i}guez \& Breger (2001) and Handler \& Shobbrook (2002), we find that a large part of the {\it Kepler} stars are concentrated near the overlap of the two instability strips, and that many members of the three groups coincide in the same region in the (\Teff, \logg)-diagram.  It will be interesting to investigate why stars with similar values of \Teff\,and \logg\,in some cases pulsate as a \dsct\,star, and in others as a \gdor\,star, or as both. Another interesting and puzzling result is that \gdor\, and \dsct\,pulsations seem to be excited in a far wider range of temperatures then previously expected. 

The distribution in \logg\,is similar for all classes. Most stars have \logg\,values between 3.5\,dex and 4.3\,dex, with  a peak around \logg $= 3.9$\,dex. We point out that the \logg\,values derived from the KIC for A-F type stars are known to have large uncertainties, and only few stars have measurements from other sources. Without more stars with accurate values derived from ground-based observations we cannot draw any conclusions.

The distribution in {\it Kepler} magnitude Kp (bottom left, Fig.\,\ref{750stars}) is representative for the distribution in Kp for \gdor\,and \dsct\,stars. It illustrates that the cut-off magnitude for the detection of \gdor\,and \dsct\,type of variations with {\it Kepler} lies around Kp\,$= 14$\,mag. The majority of the sample stars have magnitudes in the range  Kp\,$= 10-12$\,mag. 

\vsini~values are available for 41 stars of the subsample consisting of \dsct, \gdor\,and hybrid stars (see Table\,\ref{tableTeff}).  Of the five \gdor\,stars, four have  \vsini~values above 90\kms, and one has \vsini $= 15$\,\kms. Of the sixteen \dsct\,stars, eight stars have high \vsini~values, six have  moderate values (40$<$\vsini$<$90\kms), and two low values (\vsini$<$40\kms). Of the 20 hybrid stars almost all have high \vsini~values, with six stars having \vsini~values above 200\kms. Extrapolating these numbers to the full sample, we expect that many stars in the sample are moderate-to-fast rotators.

\subsubsection{Frequencies and amplitudes}
Up to 500 non-combination frequencies are detected in the {\it Kepler} time series of a single star (see Table\,\ref{classification}). These large numbers of frequencies are in sharp contrast with the small number of frequencies observed from the ground, e.g. up to 79 pulsation and combination frequencies for the \dsct\,star FG\,Vir (e.g. Breger et al. 2005) and up to 10 frequencies in the \gdor\,hybrid candidate HD\,49434 (Uytterhoeven et al. 2008), but are commonly seen in space observations because of their higher precision and sensitivity to low-amplitude variations (e.g. Poretti et al. 2009, Garc\'{\i}a Hern\'andez et al. 2009, Chapellier et al. 2011). However, it needs to be carefully checked whether all of the apparent individual frequencies are of pulsational origin.

\begin{figure*}
\centering
\begin{tabular}{ccc}\setlength{\tabcolsep}{0.1pt}
\hspace{-4mm}\resizebox{0.33\linewidth}{!}{\includegraphics{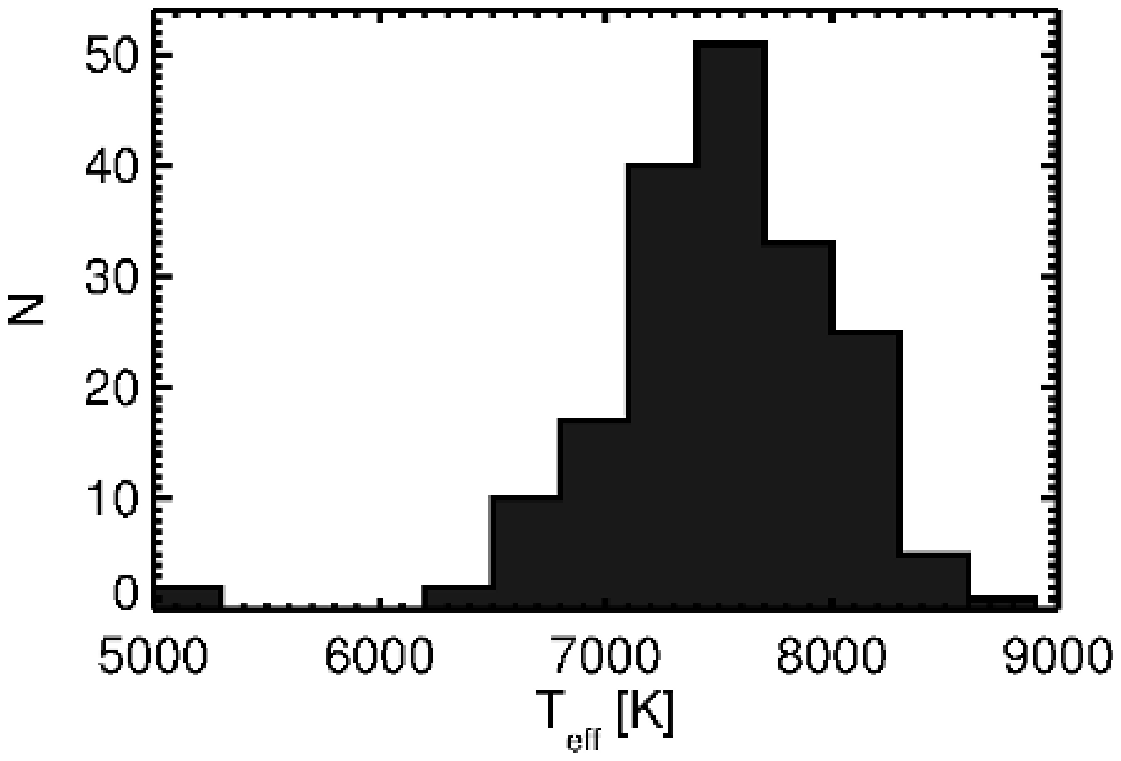}}&
\hspace{-4mm}\resizebox{0.33\linewidth}{!}{\includegraphics{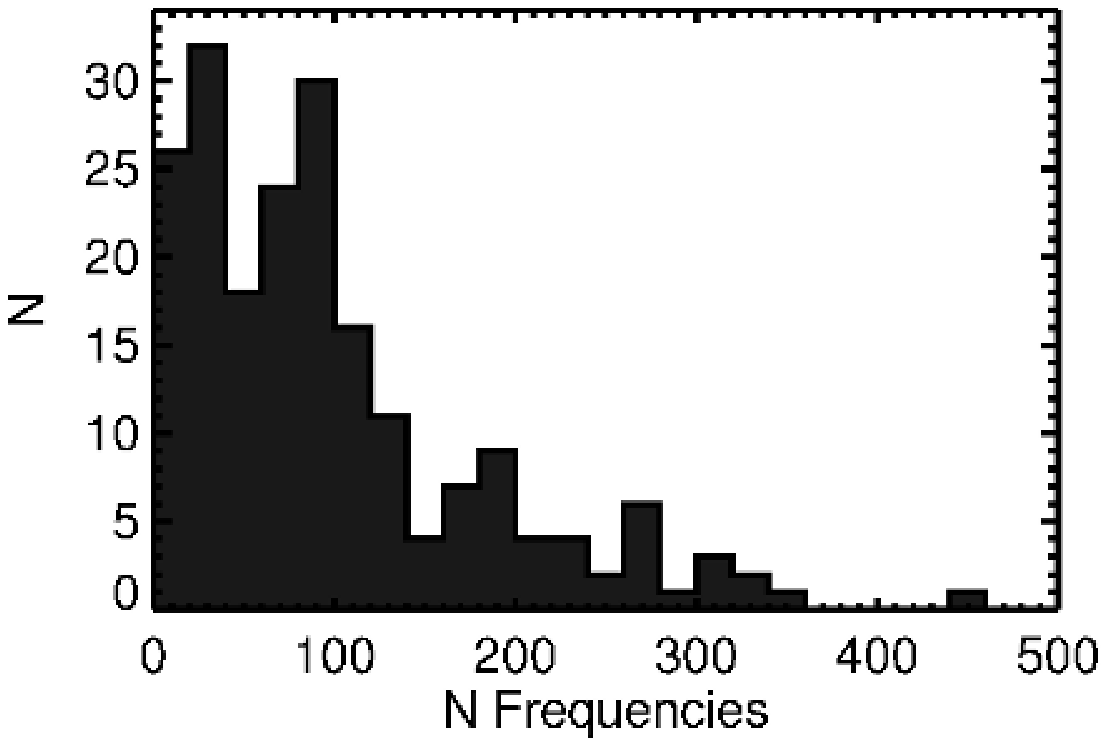}}&
\hspace{-4mm}\resizebox{0.33\linewidth}{!}{\includegraphics{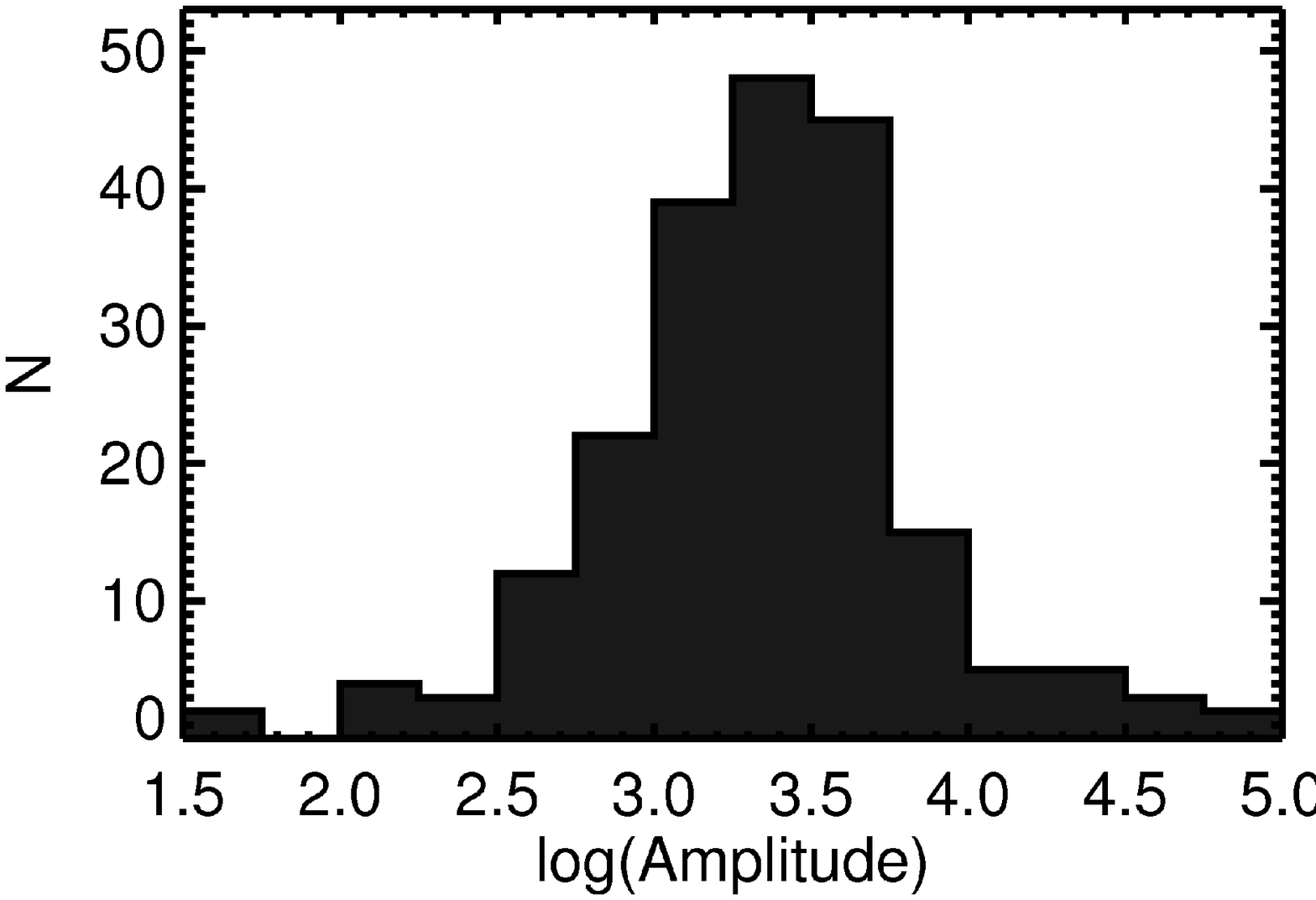}}\\
\hspace{-4mm}\resizebox{0.33\linewidth}{!}{\includegraphics{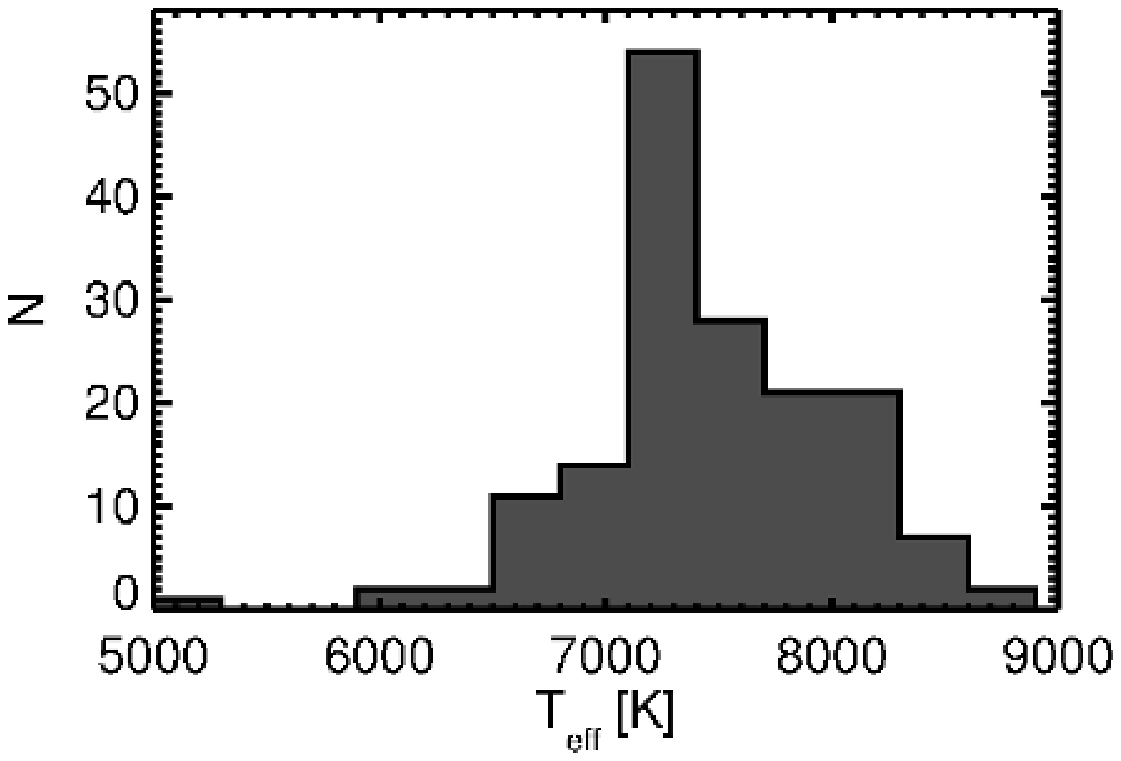}}&
\hspace{-4mm}\resizebox{0.33\linewidth}{!}{\includegraphics{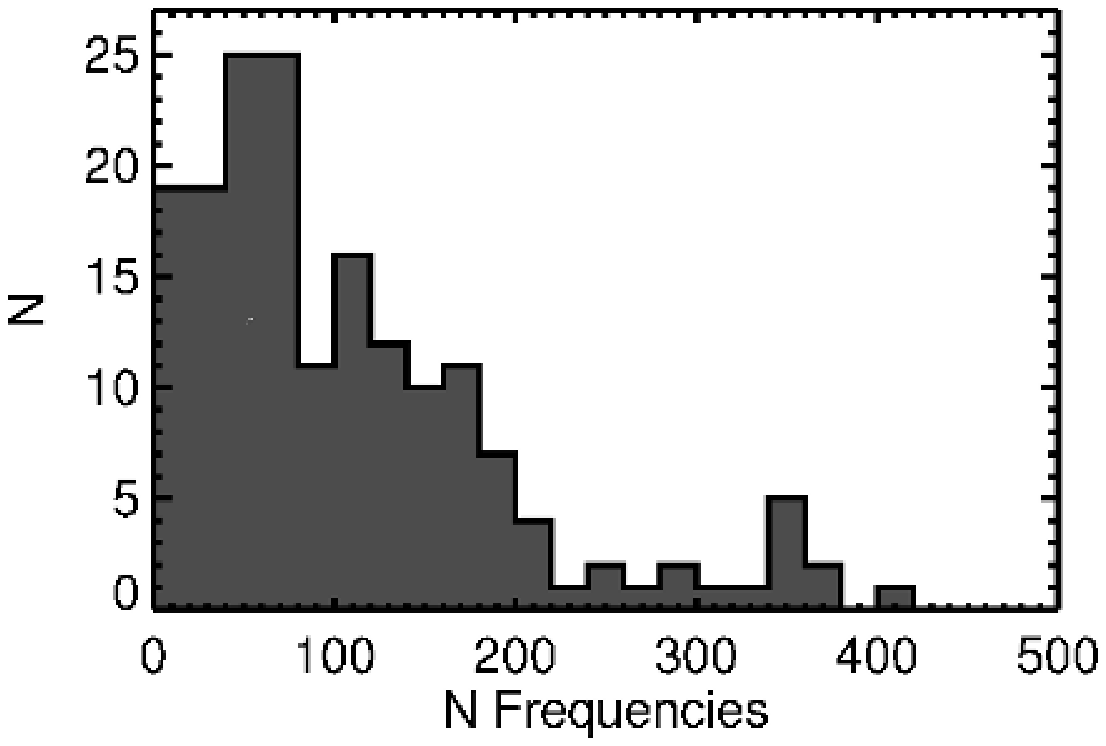}}&
\hspace{-4mm}\resizebox{0.33\linewidth}{!}{\includegraphics{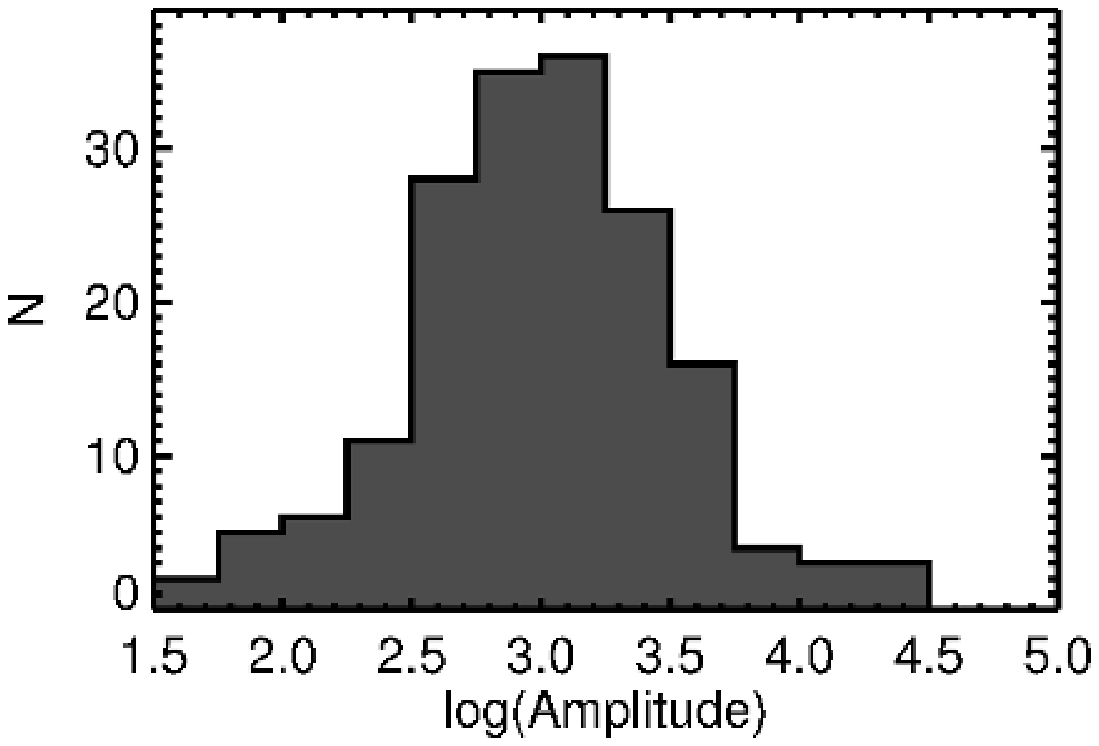}}
\\
\hspace{-4mm}\resizebox{0.33\linewidth}{!}{\includegraphics{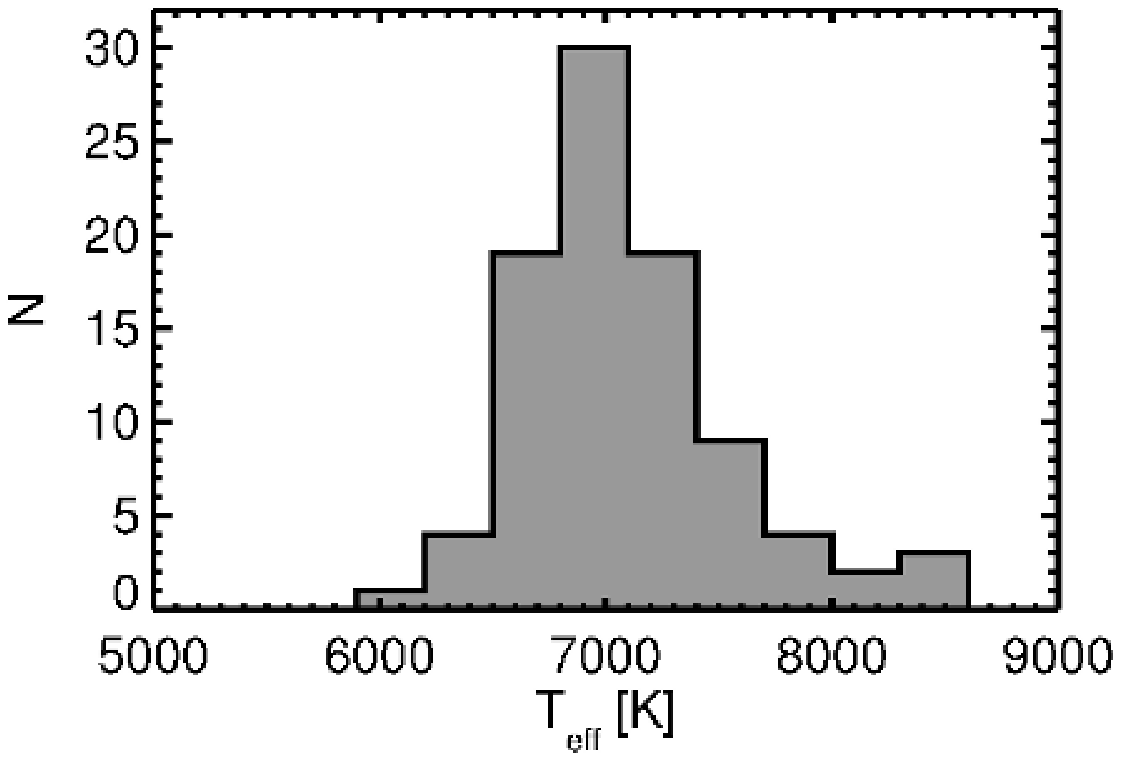}}&
\hspace{-4mm}\resizebox{0.33\linewidth}{!}{\includegraphics{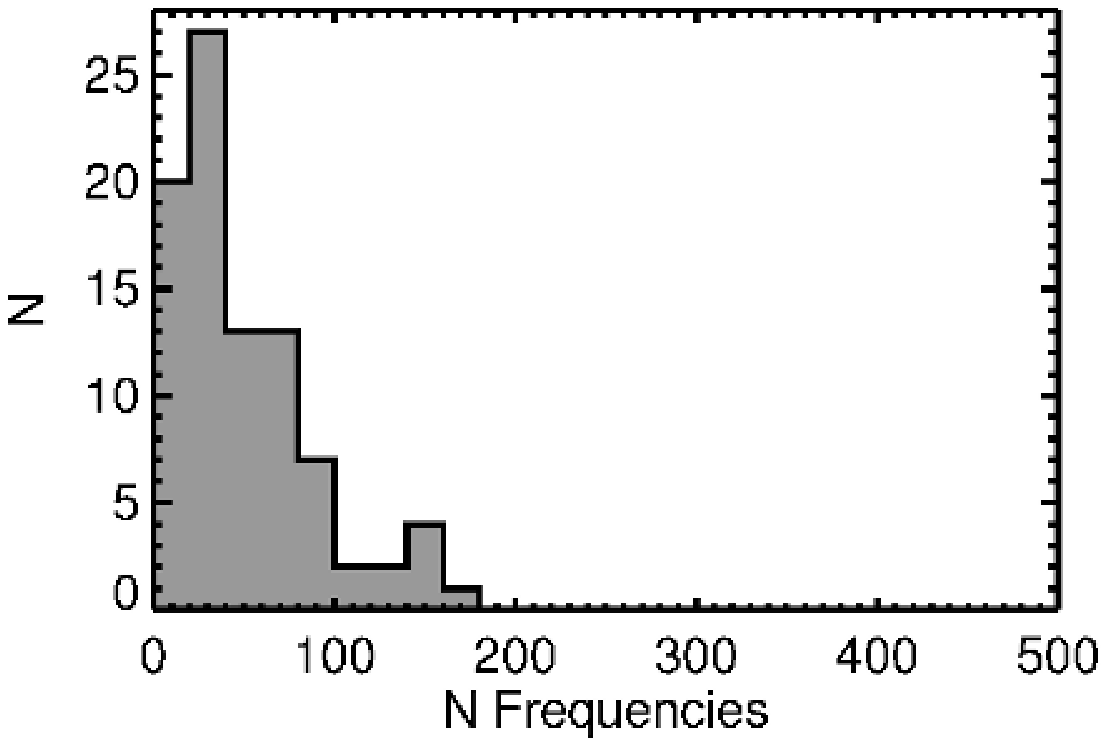}} &
\hspace{-4mm}\resizebox{0.33\linewidth}{!}{\includegraphics{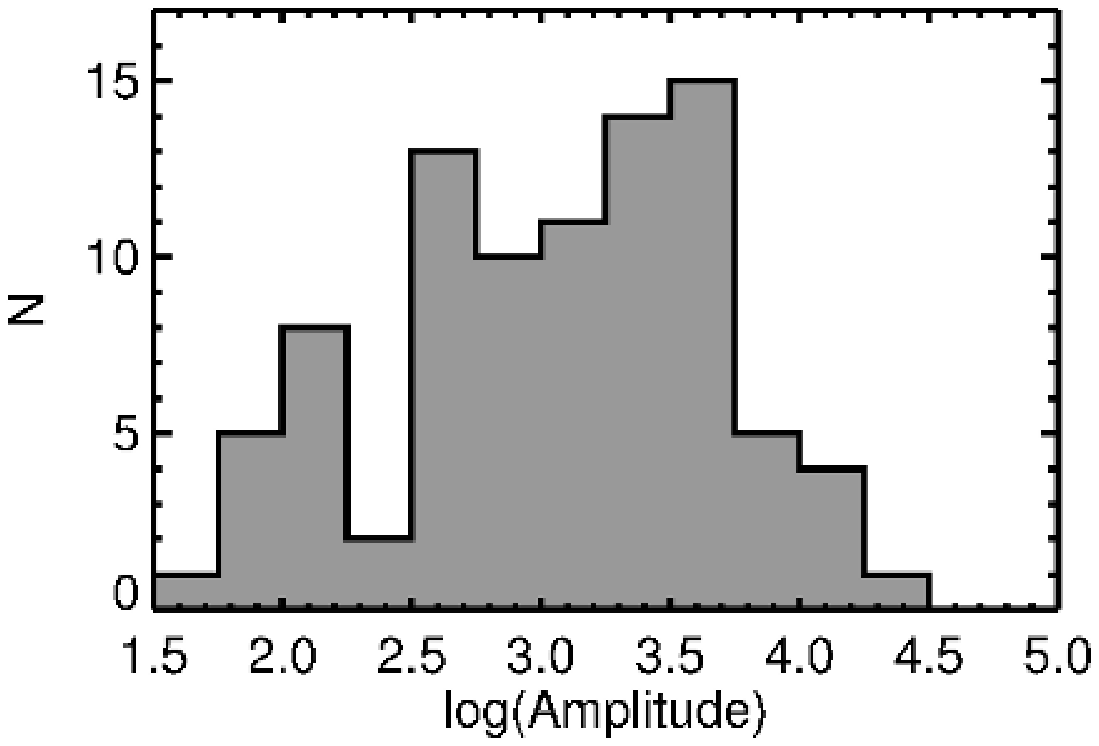}} \\
\end{tabular}
\caption{Distribution  in \Teff\,(left column),  number of detected (independent) frequencies (middle column) and  highest amplitude (ppm, in logarithmic scale) (right column), for the three groups of A-F type stars: \dsct\,stars (top; dark grey), hybrid stars (middle; middle grey), and \gdor\,stars (bottom; light grey). The number of stars belonging to each bin (N) is indicated on the Y-axis.}
\label{histogram1}
\end{figure*}

For the majority of stars (66\%), less than 100 frequencies were found, and 10\% of the stars show variations with more than 200 frequencies. If we look at the extreme cases we find that for 29 stars (6\%) fewer than 10 frequencies were detected, while for 5 stars (1\%) more than 400 frequencies were found. The middle panel of  Fig.\,\ref{histogram1} shows the distribution of the number of detected frequencies for the \dsct\,(top, dark grey),  hybrid (middle, middle grey),  and \gdor\,(bottom, light grey) stars. The highest number of frequencies are found for hybrid stars. It is worth mentioning that the number of detected frequencies versus \Teff\,follows a distribution that peaks near 7700\,K, 7500\,K, and 7000\,K for \dsct, hybrid, and \gdor\,stars, respectively. More modes are excited near the centre of the \dsct\,instability strip. For  the hybrid and \gdor\,stars most detected frequencies are found  towards the red edge of the (overlap in the) instability strip.  

The right panel of Fig.\,\ref{histogram1} shows the distribution of the highest measured amplitude in ppm logarithmic scale (log(Amplitude)) for the different groups using the same colour-code as before. The range in highest amplitude measured is 40 to 155\,000\,ppm. For about 59\% of the stars the highest amplitude is lower than 2000\,ppm.  Only 16 stars (3.5\%) show variability with highest amplitudes below 100 ppm, while 26 stars (5\%) have amplitudes above 10\,000\,ppm. In general, higher amplitudes are detected in \dsct\,pulsators than in \gdor\,stars. We point out that the origin of high peaks detected in \gdor\,stars, e.g. the amplitude of 23\,000\,ppm in the star KIC\,7304385,  is most likely related to the rotation of the star. It is worth mentioning that  amplitudes above 10\,000\,ppm are also detected in faint targets.  The highest amplitudes are found for stars within the temperature range $T_{\rm eff} = 6600-7100$\,K, which is the cool part of the instability strips.

We detected \dsct\,frequencies between 4 and 80\,d$^{-1}$. We found indications that a handful of stars vary with even shorter periods. However, these short periods need to be confirmed by means of a careful investigation of the specific frequency spectra, which is beyond the scope of this paper.

When considering the \dsct\,stars and hybrid classes, which amount to atotal of 375 stars, we find that  56\% shows an upper frequency limit between 40 and 70\,d$^{-1}$. Only 10\% of the \dsct\,and hybrid stars have frequencies up to 80\,d$^{-1}$, and 9\% only show variations with frequencies lower than 20\,d$^{-1}$. We note that \gdor-dominated hybrids that show variations with frequencies  higher than 60\,d$^{-1}$ are rare (three stars in our sample).

The majority of the hybrid stars detected in the {\it Kepler} data show all kinds of periodicities within the \gdor\,and \dsct\,range (see columns 6 and 7 in Table\,\ref{classification} which give the frequency range of the detected frequencies in the \gdor\,and \dsct\,domains). This observational fact is interesting because from a theoretical point of view no excited modes are expected between about 5 and 10\,d$^{-1}$, i.e. the so-called 'frequency gap' (see, e.g. Grigahc\`ene et al. 2010).  Only for five hybrid stars a 'frequency gap' is observed\footnote{The six hybrid stars that show a 'frequency gap' are: KIC\,3851151, KIC\,4556345,  KIC\,7770282, KIC\,9052363, and KIC\,9775454.}. Possible explanations for the absence of gaps, within the present non-adiabatic theories, are  that the frequencies within the gap are high-degree and/or rotationally split modes  (Bouabid et al. 2009). 

\section{A first step towards understanding the relation between \dsct, \gdor, and hybrid stars}
\label{Sectenergy}
As presented in the previous section,  it is not trivial to distinguish between the three groups of variable A-F type stars defined in Section\,\ref{sectclass}. The relation between the three groups is currently unclear as well because \dsct, \gdor, and hybrid stars coincide in the (\Teff, \logg)-diagram (Fig.\,\ref{logTefflogg}). Driven by the idea to find observables based on physical concepts that allow insight in the different internal physics of the three types of stars, we constructed two new observables that can provide an alternative way to improve our understanding of the relation between the three groups. We point out that several observational parameters can be found that reflect the different inherent properties of the three groups  in one way or another. For instance, \dsct\,stars pulsate with shorter periods, and are generally hotter than \gdor\,stars. A combination of these parameters  will lead to a differentiation of  the groups, such as for instance a (\Teff, $f_{\rm max}$)-diagram, with $f_{\rm max}$ the frequency associated to the highest amplitude mode.  However, we emphasize that our aim is to find observables that can be directly related to the internal physics of the stars.

According to the current instability theories, which need to be revised following the results presented in this work, the main driving process of the oscillations in \dsct\,stars is related to the opacity variations in the ionization zones (Unno et al. 1989). These zones are located in the region where the main energy transport mechanism is convection and where a small quantity of energy is transported by radiation. The total amount of driving energy going into the mode is directly related to the radiative luminosity in this zone, and this latter quantity is a function of the convective efficiency. Therefore, we expect a relation between the energy of the observed modes and the convective efficiency of the outer convective zone. We searched for this relation and constructed two observables, {\it energy} and {\it efficiency}, that are estimates of the energy and the convective efficiency, using the available observational data.

\subsection{Energy}
The kinetic energy of a wave is given by
\begin{equation}
E_{kin}={1\over 2}f(\rho_*)(A\zeta)^2,
\end{equation}
where $f$ is a function of the stellar density $\rho_*$, $A$ is the amplitude of the
oscillation, and $\zeta$ is the pulsation frequency. Using the available observational data, we construct the following observable that we call {\it energy}, which is a first approximation and estimate of the kinetic energy of the wave:
\begin{equation}
energy \equiv (A_{\rm max} \zeta_{\rm max})^2,
\end{equation}
where $A_{\rm max}$ and $\zeta_{\rm max}$ refer to the highest amplitude mode of the star (in ppm), and associated frequency (in d$^{-1}$). The pulsation amplitude is a function of the observed amplitude and the relative variation of the flux, and is given by the expression (Moya \&  Rodr{\'{\i}}guez-L{\'o}pez 2010):
\begin{equation}
{\Delta R/R}=-{\Delta m\over \ln (5+10 dT)},
\end{equation}
where $\Delta R\over R$ is the relative pulsational amplitude, $\Delta m$
 the observed magnitude variation of the mode, and $dT$ is given by
\begin{equation}
dT={\delta T_{\rm eff}\over T_{\rm eff}}/{\xi_r\over r},\,\,(r=R),
\end{equation}
with $\xi_r$ the variation in radius of the mode, and $dT$ is evaluated at the surface of the star ($r=R$).

\begin{figure}
\centering
\begin{tabular}{cc}
\setlength{\tabcolsep}{0.1pt}
\hspace{-4mm}\resizebox{0.44\linewidth}{!}{\includegraphics{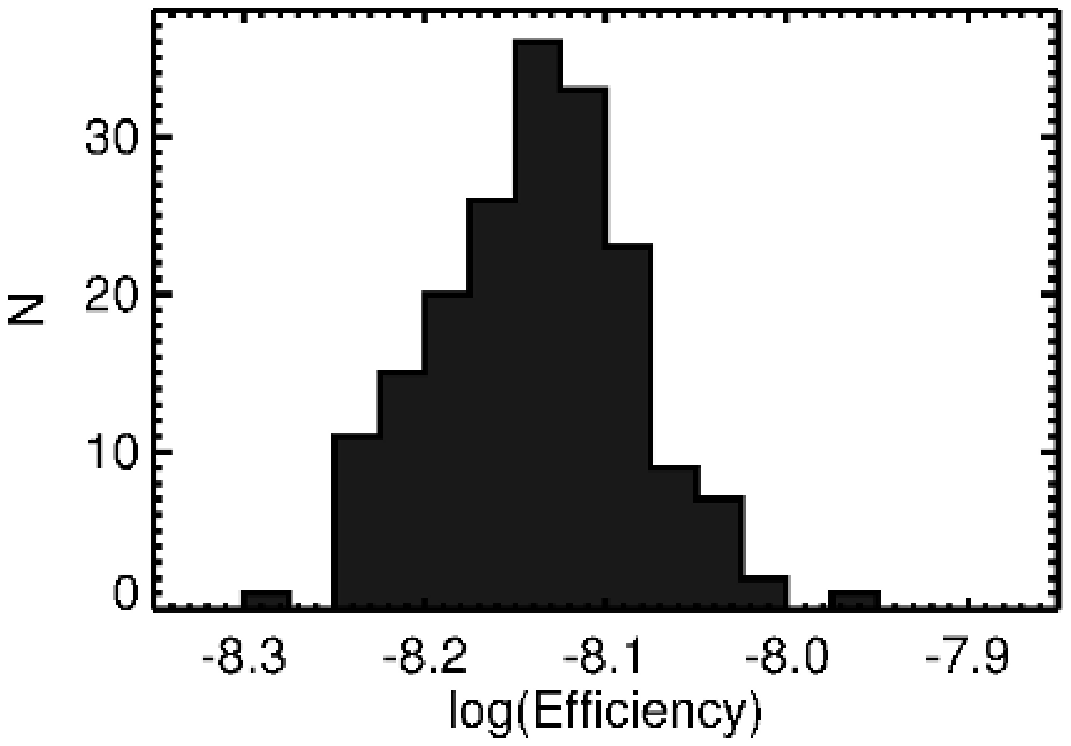}}
&
\hspace{-4mm}\resizebox{0.44\linewidth}{!}{\includegraphics{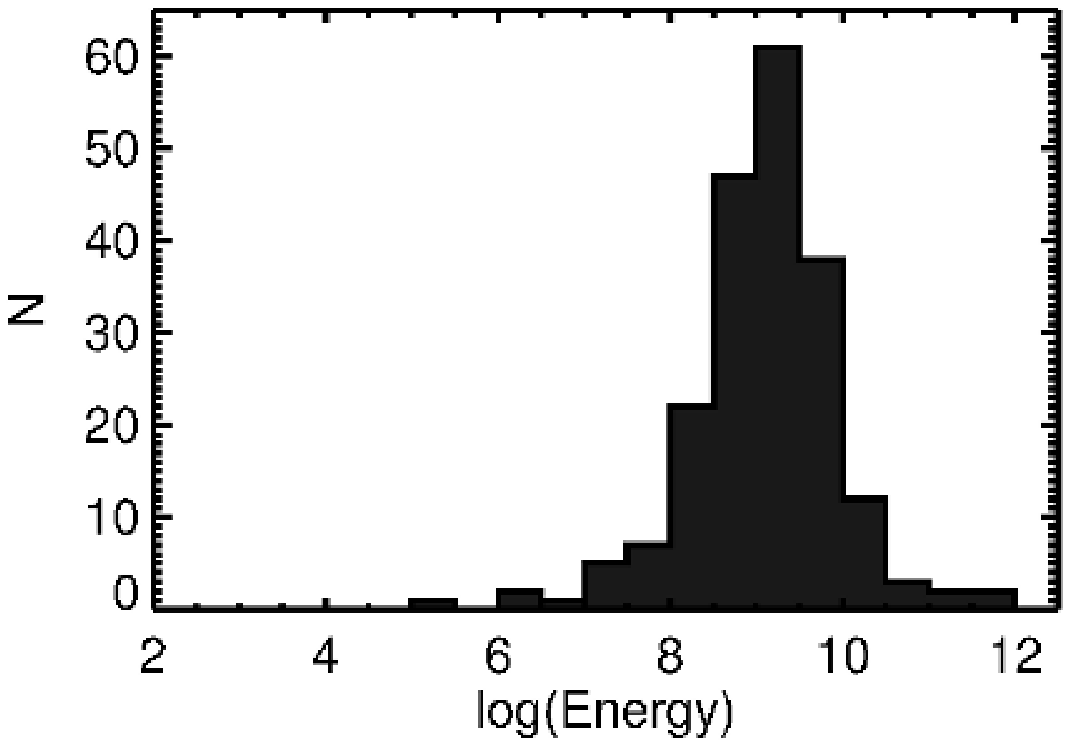}}
\\
\hspace{-4mm}\resizebox{0.44\linewidth}{!}{\includegraphics{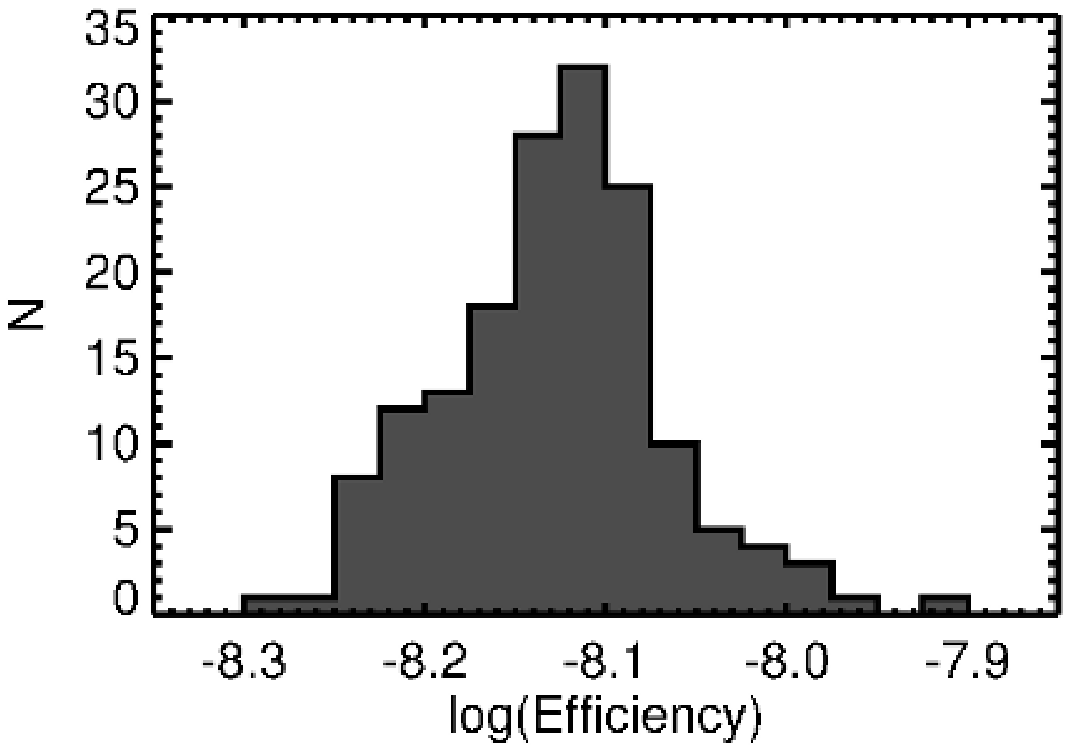}}
&
\hspace{-4mm}\resizebox{0.44\linewidth}{!}{\includegraphics{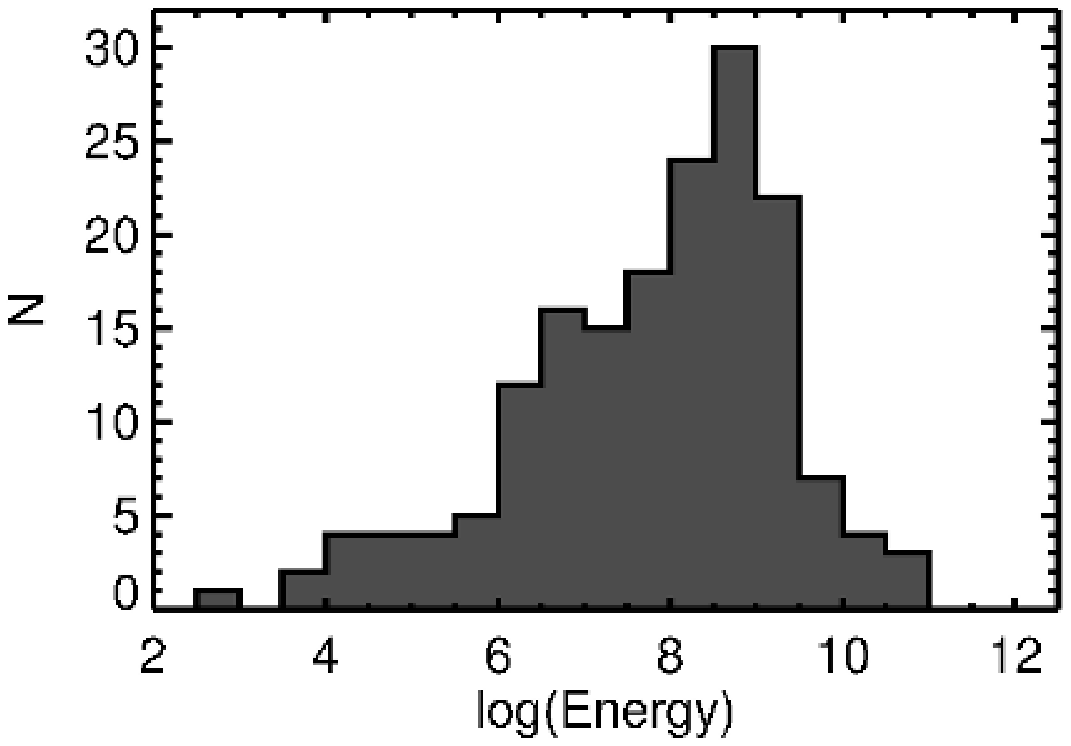}}
\\
\hspace{-4mm}\resizebox{0.44\linewidth}{!}{\includegraphics{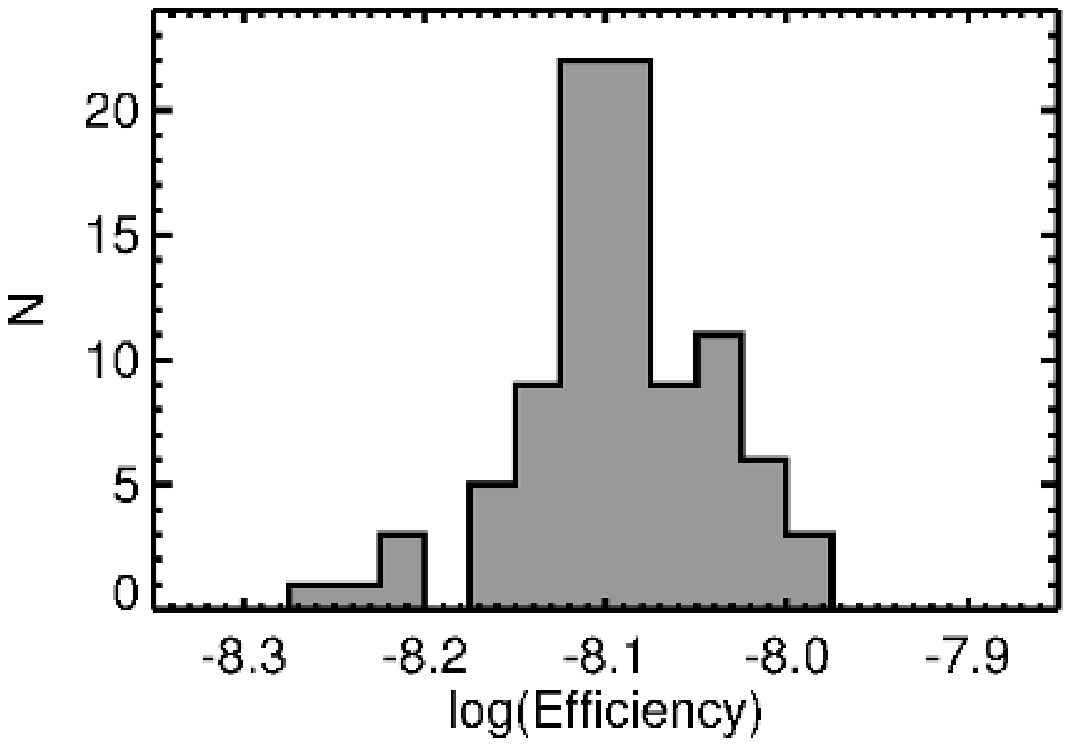}}
&
\hspace{-4mm}\resizebox{0.44\linewidth}{!}{\includegraphics{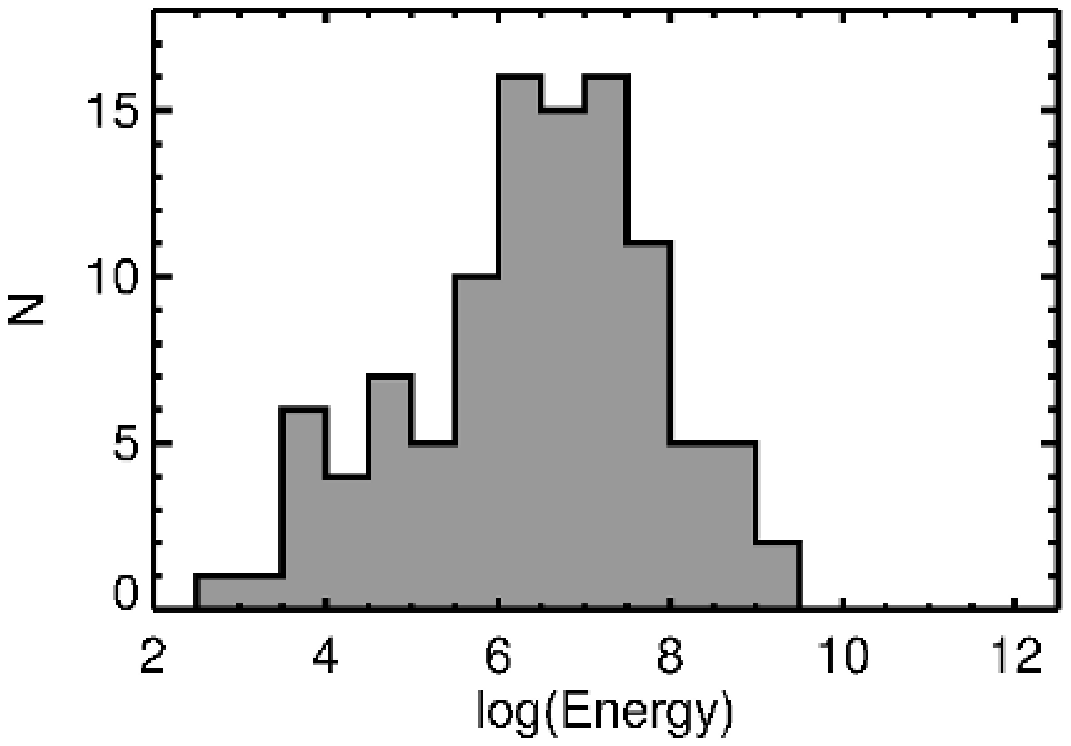}}
\\
\end{tabular}
\caption{Distribution  in $\log$({\it energy}) (right),  and $\log$({\it efficiency}) (left) for the \dsct\,(top, dark grey),  hybrid (middle, middle grey), and \gdor\,(bottom, light grey) stars. The number of stars belonging to each bin (N) is indicated on the Y-axis.}
\label{enef}
\end{figure}

Non-adiabatic calculations of a representative model of a hybrid
pulsating AF-type star including time dependent convection (Grigahc{\`e}ne et al. 2005) show that the difference between the predicted $dT$ value of
asymptotic g-modes (\gdor\,stars) and low-order p-modes (\dsct\,stars) is around one order of magnitude or less, where the
\dsct\,stars have higher values. Therefore, we can directly use the observed magnitude variation as a measurement of the radial amplitude
variation. That we are using an approximation does not change the conclusions of the present study, because the observed differences are larger than two
orders of magnitude (see Figs\,\ref{enef} and \ref{diagram}).

The right column of Fig.\,\ref{enef} shows the distribution in $\log$({\it energy}) for the \dsct\,(top, dark grey), hybrid (middle, middle grey), and \gdor\,(bottom, light grey) stars. Clearly,  the weight of the distribution is located in the region $\log$({\it energy})$> 8$ for stars dominated by frequencies in the \dsct\,domain, and in the region $\log$({\it energy})$<8$ for stars with dominant \gdor\,pulsations.

\subsection{Efficiency}
In the introduction of this section we pointed out that a relation between the convective efficiency and mode excitation can exist. Recent studies on convective efficiency of the outer convective zone of F-G-K stars using 3D models show that the convective efficiency is related to the position of the star in the Hertzsprung-Russell (HR-) diagram (Trampedach \& Stein 2011). 
To construct an observable related to the convective efficiency that can be described with only variables related with the position in the HR-diagram, we found inspiration in the analytic description of the convective energy given by the mixing length theory\footnote{In analogy to the mean free parameter in gas kinetic theory, the mixing length is defined as the mean distance over which a fluid bulb conserves its properties. Generally, the mixing-length is assumed to be proportional to the  pressure-scale height by a factor $\alpha$ that is usually called mixing-length parameter.} (B{\"o}hm-Vitense 1958).
There, the convective efficiency, $\Gamma$, is defined as
\begin{equation}
\Gamma = \left [ \frac{A^2}{a_0} (\bigtriangledown_{\rm rad} - \bigtriangledown) \right ]^{1/3},  \label{convective}
\end{equation}
 with $a_0$ a constant, $\bigtriangledown_{\rm rad}$ and $\bigtriangledown$ the radiative and real temperature gradient, respectively, and  
\begin{equation}
A \sim {c_p\kappa p \rho c_s \alpha^2 \over 9 \sigma T^3 g \sqrt{2 \Gamma_1}}
\end{equation}
 (see Cox \& Giuli 1968).

This quantity, which measures the ratio between the convective and radiative conductivity, depends on a large number of physical variables: the specific heat capacity at constant pressure $c_p$, the opacity $\kappa$, the pressure $p$, the stellar density $\rho$, the sound velocity $c_s$, the mixing length parameter $\alpha$, the Stephan-Boltzmann constant $\sigma$, the temperature $T$, the gravity $g$, and the first adiabatic coefficient $\Gamma_1$. Because we only have information on a limited number of observational variables, our estimate of the quantity is only an approximation. Inspired by these equations, we searched for the combination of temperature and gravity that empirically provided the best means to separate between \gdor\,and \dsct\,stars (see statistical test below), and define the observable ${\it efficiency}$ as
\begin{equation}
{\it efficiency}\equiv (T_{\rm eff}^3\log g)^{-2/3} \sim \Gamma.
\end{equation}

Because the efficiency of the convective zone is expected to be higher for \gdor\,stars than for \dsct\,stars, the observable {\it efficiency} should have a  higher value for  \gdor\,stars than \dsct\,stars. This behaviour is indeed observed, as illustrated in the left panel of Fig.\,\ref{enef}, where the distribution in $\log$({\it efficiency}) is given. The majority of \dsct\,stars have values $\log$({\it efficiency})$ <-8.1$\,dex, while the histograms for \gdor\,pulsators peak in the region $\log$({\it efficiency})$ > -8.1$\,dex. 

\subsection{{\it Efficiency} versus {\it energy}}
\begin{figure*}
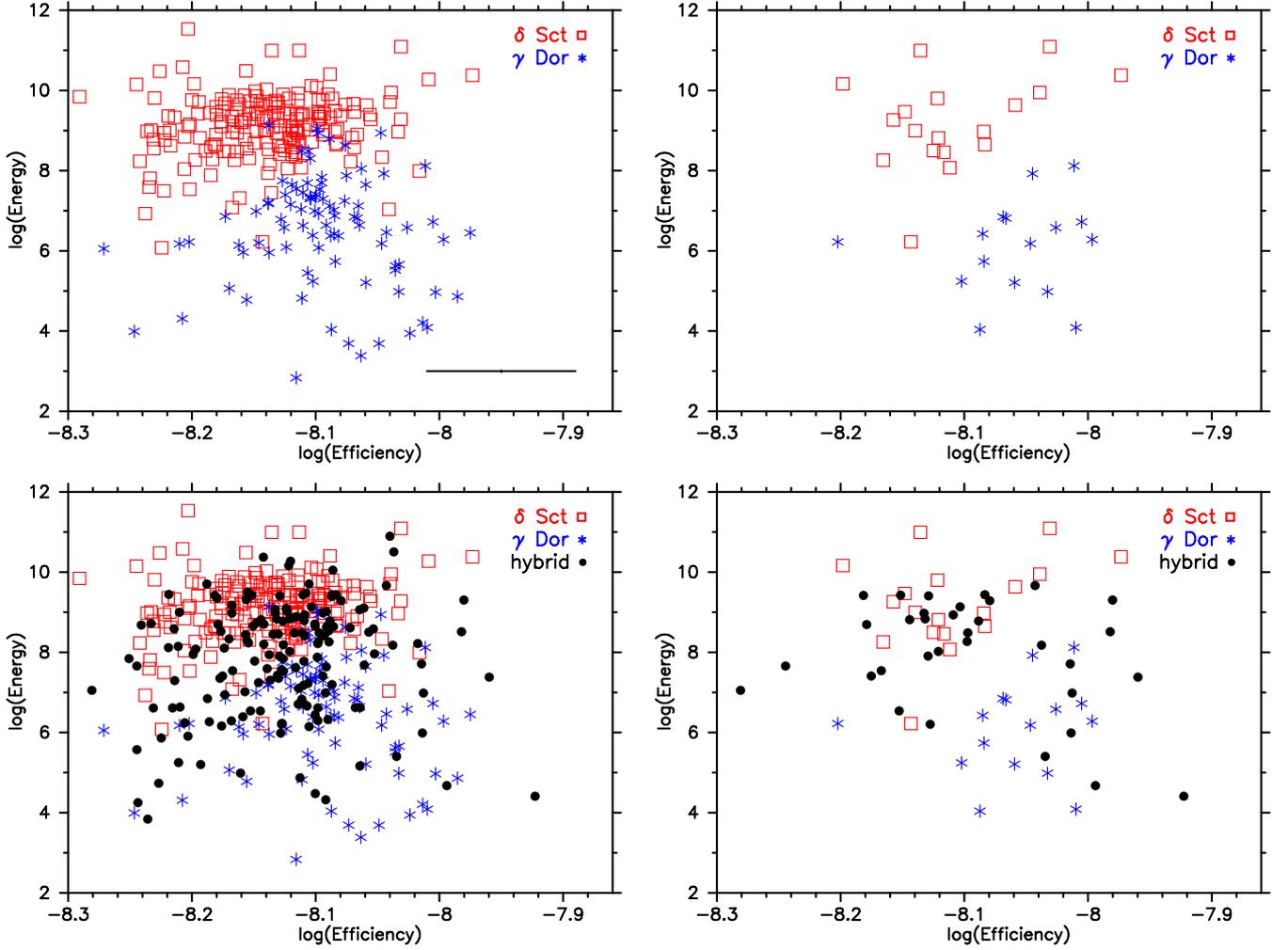

\centering
\begin{tabular}{cc}
\resizebox{0.47\linewidth}{!}{\rotatebox{-90}{\includegraphics{17368fig35.eps}}}&
\resizebox{0.47\linewidth}{!}{\rotatebox{-90}{\includegraphics{17368fig36.eps}}}\\
\resizebox{0.47\linewidth}{!}{\rotatebox{-90}{\includegraphics{17368fig37.eps}}} &
\resizebox{0.47\linewidth}{!}{\rotatebox{-90}{\includegraphics{17368fig38.eps}}} \\
\end{tabular}
\caption{Observable $\log$({\it energy}) plotted versus $\log$({\it efficiency}) for  \dsct\,(open squares) and \gdor\,(asterisks) stars only (top), and for hybrid stars as well (bullets) (bottom). The left panels include all 480 {\it Kepler} stars that are assigned to one of the three groups. The right panels show the 49 stars for which reliable values of \Teff\,and \logg\,are available. In the on-line version of the paper the symbols representing the \dsct, \gdor\, and hybrid stars are red,  blue, and  black, respectively. The cross in the right bottom corner of the top left panel represents the typical error bars on the values:  0.04\,dex and 0.12\,dex for $\log$({\it energy}) and $\log$({\it efficiency}), respectively.}
\label{diagram}
\end{figure*}

When we plot the two new observables, $\log$({\it energy}) versus $\log$({\it efficiency}), the groups of \dsct\, and \gdor\,stars are fairly well separated (see top panels Fig.\,\ref{diagram}). A  $\log$({\it energy}) value of 8 leaves 90\% of the \dsct\,and \gdor\,stars separated. The  bottom panel of Fig.\,\ref{diagram} shows the same diagram with values for the hybrid stars included, using the same colours and symbols as before. Typical errors on the values are 0.04\,dex and 0.12\,dex for $\log$({\it energy}) and $\log$({\it efficiency}), respectively. The hybrid stars  are placed in the intermediate region. We observed that \dsct\,(\gdor) dominated hybrids fall in the same region as the \dsct\,(\gdor) stars.  

We performed a Mann-Whitney U test with an adapted p-value (p = 0.0166) according to the closed-test principle described in  Horn \& Vollandt (1995), to statistically investigate if the mean of the distribution in $\log$({\it energy}) and $\log$({\it efficiency}) is different for the three different groups. The test shows that the difference  in the mean of the distributions in both $\log$({\it energy}) and $\log$({\it efficiency}) is statistically significant for all groups. However, the apparent separation in  $\log$({\it efficiency}) becomes less evident when we take the considerable error bars into account. We also performed a $\chi^2$ test (as described by Press et al. 1992) to determine if the distributions
themselves were different. All distributions are statistically significant, save for the \gdor\,versus hybrid star {\it efficiencies}, where they are marginally similar. This conclusion
holds even if we vary the \Teff\,and \logg\,values within the error ranges and recompute the {\it efficiencies} or vary the inputs into the {\it energies}.  We point out once again that the definition of {\it efficiency} is only a rough estimate of the theoretical expression for the convective efficiency, and might - at this stage - not be refined enough to display the separating power between the groups we expect the convective efficiency to have. In a follow-up investigation we will assess the goodness of approximation of our definition of {\it efficiency} by comparison with values of the convective efficiency as given by Eq.~(\ref{convective}), calculated for several model stars, and finetune its definition.

The two new approximate observables {\it energy} and {\it efficiency} reflect the different internal physics of  oscillators with dominant \dsct\,pulsations and oscillators dominated by \gdor\,pulsations, and seem to allow us to distinguish between them. However, it needs to be further investigated if the two observables can be considered as independent parameters. This, together with an exploration of the  physical mechanisms behind the instability of these stars, is the topic of a forthcoming paper. The observables {\it energy} and {\it efficiency} are promising starting points to explore the relation between \dsct, \gdor\,and hybrid stars, but need to be refined.

\section{Summary, discussion, and future prospects}
We analysed the {\it Kepler} light curves based on survey phase data with time spans between 9\,d and 322\,d available through KASOC and associated frequency spectra of 750 candidate A-F type stars in search for \dsct, \gdor, and hybrid pulsators. The main results are:

\begin{itemize}
\item The {\it Kepler} light curves of the sample of 750 candidate A-F type stars show a variety in variability behaviour.
\item Observationally, we propose three main groups to describe the observed variety:  \gdor, \dsct, and hybrid stars. The latter group includes both \dsct-dominated and \gdor-dominated hybrid stars. About 63\% of the sample are unambiguously assigned to one of the three groups. 
\item About 23\% of the sample are hybrid candidates (171 stars, or 36\% of the stars assigned to the three groups). This is in strong contrast with the number of hybrid candidates so far observed from the ground, but compatible with the first {\it Kepler} study of \gdor\,and \dsct\,variables by Grigahc\`ene et al. (2010).  The far superior precision of the {\it Kepler} space data opens a new window in detecting low-amplitude variations. {\it Kepler} will be ideal to study hybrid behaviour in different types of stars, such as roAp stars (Balona et al. 2011a), sdB stars (\O{}stensen et al. 2010), and B stars (Balona et al. 2011b).
\item We presented a characterization of the stars in terms of number of detected frequencies, frequency range, and typical pulsation amplitudes, which provides valuable feedback for models and instability studies. This is the first time that this kind of information is available for a substantial sample of stars. Up to 500 non-combination frequencies are detected in the {\it Kepler} time series of a single star. The highest pulsation amplitude measured is 58\,000\,ppm. The shortest detected \dsct\,periods are about 18\,min. We find that hybrid stars show all kinds of periodicities within the \gdor\,and \dsct\,range. In particular, the majority of hybrid stars shows frequencies between 5 and  10\,d$^{-1}$. From a theoretical point of view, this result presents a number of challenges, because the currently accepted over-stability mechanisms cannot explain the presence of pulsational modes in the wide frequency ranges observed with {\it Kepler}. It needs to be investigated if and to what extent the presence of stochastic modes, high-degree, and/or rotationally split modes with high amplitudes, granulation and effects of convection can explain part of the unexpected observed modes. 
\item The location of \gdor\,and \dsct\,classes in the (\Teff, \logg)-diagram has been extended (Fig.\,\ref{logTefflogg}).  We find indications that {\it Kepler} \dsct\,stars exist beyond the red edge of the observational instability strip, while {\it Kepler} \gdor\,pulsations seem to appear in both hotter and cooler stars than observed so far. The {\it Kepler} hybrid stars occupy the entire region between the blue edge of the \dsct\,instability strip and the red edge of the \gdor\,instability strip and beyond.  These results, if confirmed by verification of the temperature and \logg\,values in a more comprehensive sample, imply that the observational instability strips need to be extended to accommodate the {\it Kepler} \dsct\,and \gdor\,stars. From a theoretical point of view, the overall presence of hybrid stars implies an investigation of other pulsation mechanisms to supplement the $\kappa$\,mechanism and convective blocking effect to drive hybrid pulsations. 
\item Two new 'observables' that reflect the different internal physics of \dsct\,and \gdor\,pulsators are introduced to investigate the relation between the two types of pulsations (Fig.\,\ref{diagram}): (1)  {\it efficiency},  related to the convective efficiency of the outer convective zone, and a function of \Teff\,and \logg;  and (2) {\it energy}, the driving energy of a mode, and a function of the highest observed frequency amplitude and the associated frequency. Both observables are empirical and are constructed using only available measured variables. The impact and physical significance of the group separation in the ($\log$({\it efficiency}), $\log$({\it energy}))-diagram needs to be  investigated in more detail. The two new observables are a promising starting point for further investigations of the relation between \dsct, \gdor\,and hybrid stars.
\item Our study indicates that Kp\,$= 14$\,mag is a cut-off magnitude for detection of variations with amplitudes below 20\,ppm in A-F type stars with {\it Kepler}. 
\item Sixteen percent of the sample stars show no clear variability within the expected range of frequencies for \dsct\,and \gdor\,stars. Faint and cool stars predominate this sample. Among the stars, we identified 75 candidate solar-like stars. No correlation between  non-variability and the length of the available dataset or the available cadence mode is found. We find indications for the presence of constant stars inside the instability strips of A-F type pulsators.
\item The remaining 21\% of sample stars are identified as a Cepheid, B-type stars, red giant stars, stars that show stellar activity, or binaries. At least 12\% of the sample are identified as a binary or multiple system, based on investigation of the {\it Kepler} light curve or on input from the literature. Many long-period binaries are expected to be among the remaining stars of the sample.  3.5\% of the sample stars shows eclipses. Several of the EBs have variable components, including  \dsct, \gdor, and hybrid stars. 
\end{itemize}

Clearly, space missions are changing the landscape of  \gdor\,and \dsct\,pulsators. We aimed at a global analysis of the sample stars. A careful seismic analysis of individual stars is needed to confirm their classification,  clarify the observed variety in pulsational behaviour, fully characterize the properties of the \dsct, \gdor, and hybrid groups,  understand their relationship, clarify the driving mechanism(s) for each group,  and elaborate on the variables {\it energy} and {\it efficiency}. The observational results with {\it Kepler} presented here open up several new questions and theoretical challenges for the current models related to pulsational instability, thermodynamics, and stellar structure. We mention here some topics for further investigation.

To be able to place the stars confidently in the (\Teff, \logg)-diagram,  estimate the projected rotational velocity, and  derive accurate abundances, at least one high-resolution spectrum is needed for each star. To this end, an observational campaign is ongoing (Uytterhoeven et al. 2010a,b). Most stars of the \dsct, \gdor, and hybrid stars in our sample with magnitude Kp\,$\leq 10.5$\,mag have recently been observed or are scheduled to be observed in the coming months. However,  70 \% of the stars in Table\,\ref{classification} are fainter than magnitude Kp\,$= 10.5$\,mag, for which it is time-consuming and less practical to observe them with the available 2-m class telescopes that are equipped with a high-resolution spectrograph.  

Because the oscillation modes in A-F type stars do not produce evident frequency patterns in their mode spectra, as is the case for solar-like oscillators, the identification of pulsation modes benefits from high-resolution spectral or multi-colour time series. Here we encounter limitations owing to the relative faintness of the {\it Kepler} sample too. For instance, it is only feasible to efficiently spectroscopically monitor the few brightest (Kp\,$\leq 9$\,mag) stars from the ground, while multi-colour photometry can go a few magnitudes fainter. Moreover, it will be impossible with the current instrumentation to detect the pulsation amplitudes of the order of a few $\mu$mag from the ground. Therefore, only for a limited selection of the stars in Table\,\ref{classification}, i.e. bright stars exhibiting  high-amplitude variations, will it  be feasible to organize ground-based follow-up campaigns.

 For all other stars, we will have to rely on extracting information on the pulsation modes directly from frequency patterns observed in the {\it Kepler} data. Quasi-periodic patterns have been observed before in \dsct\,stars (Handler et al. 1997; Garc\'{\i}a Hern\'andez et al. 2009). But in fast rotating stars the rotation destroys regular frequency and period patterns of p- and g-modes, which complicates the mode identification (e.g.\,Ligni\`eres et al. 2006; Ballot et al. 2010).  For slowly rotating  g-mode pulsators ($V_{\rm rot} < 70$ km s$^{-1}$), a mode-identification technique has been developed that relies only on accurate values of at least three frequencies (Frequency Ratio Method, Moya et al. 2005; Su\'arez et al. 2005), which is ideal to apply to the information extracted from the {\it Kepler} white light, without colour or spectral information. Unfortunately, many of our stars are moderate-to-fast rotators (see Sect.\,\ref{sectChar}). Hence, the mode identification will be very challenging and will require  more investigation.

An individual analysis of the candidate hybrid stars is needed to confirm their hybrid status and to firmly characterize their pulsation properties. The current theoretical instability models for hybrid stars need to be revised to be able to accommodate all  stars that have been proposed as hybrid candidates in this paper. This includes a revision of the mechanisms that allow driving of p-\,and g-modes in A-F type stars with a broad range of temperatures. Additional processes that can be investigated with possible effect on the driving are stochastic excitation (Houdek et al. 1999; Samadi et al. 2002), a convective driving mechanism similar to  g-mode pulsations in white dwarfs (Goldreich \& Wu 1999),  a $\kappa$\,mechanism-related effect presented by Gautschy and L\"offler (1996) and L\"offler (2000), and radiative levitation (Turcotte et al. 2000).
Asteroseismic diagnostics have been studied to find signatures of stochastic mechanisms at the origin of the instability of \gdor\,oscillators (Pereira et al. 2007). In that work, this possibility was not discarded, but continuous and precise space data were not yet available. The {\it Kepler} time series of the sample of stars studied here will be an ideal new testbed for this method. 

The long, continuous time series that {\it Kepler} will deliver during its lifetime will unveil a large number of amplitudes at $\mu$mag level. This precision will open up opportunities to search for signatures of granulation in the variable star light (Kallinger \& Matthews 2010). Spectroscopically, convective signatures have been detected in the microturbulence and line broadening of A-F type stars cooler than \Teff $= 10,000$\,K (Landstreet et al. 2009).

   Also the theoretical instability strips of the \gdor\,and \dsct\,pulsators need revision. As shown in Fig.\,\ref{Teffloggsample}, stars exhibiting purely \gdor\,or \dsct\,pulsations seem to exist beyond the current blue and red edge of the respective instability strips. Moreover, it is worth investigating if the evolutionary phase of \gdor\,stars can be derived from properties in their frequency spectra, as is recently suggested by Bouabid et al. (2011), based on a theoretical study of seismic properties of MS and pre-MS  \gdor\,pulsators.  

Another open question is the existence of non-variable A-F type stars inside the instability strips. So far, it is suggested (Poretti et al. 2003; Breger 2004) that all seemingly constant stars in the instability strip are low-amplitude pulsators. In this study we find indications that non-variability exists within the instability strip, but a more in-depth investigation based on a more comprehensive sample of stars with precise values of \Teff\,and \logg\,is needed to confirm this.  

Furthermore, candidate \gdor\,stars with only a few excited dominant modes deserve to be looked at in more detail. The relation between rotation and pulsations is not yet clear. Moreover, the differentiation between pulsations and rotational variability proves to be very difficult (Breger 2011; Monnier et al. 2010). In the pilot study by Balona et al. (2011d) it was suggested that pulsation and rotation periods might be very closely related. It needs to be investigated to which extent the rotation influences the excitation of the observed modes. To help this investigation, \vsini~values are needed.

Constraints on important physical parameters that are crucial for seismic modelling, such as stellar radius and mass, can be derived directly for pulsators in binary systems (e.g. Tango et al. 2006; Desmet et al. 2010). Our sample consists of several binaries and eclipsing systems with (a) pulsating component(s) (see Table\,\ref{other}). Hence, these targets in particular are very promising for dedicated ground-based follow-up observations, and a seismic analysis. Moreover, it will be interesting to investigate the effect of tidal interactions on pulsation frequencies (e.g. Uytterhoeven et al. 2004; Derekas et al. 2011).

Four of the EBs with a candidate \gdor, \dsct, or hybrid component in our sample are known as chemically peculiar stars (see Table\,\ref{database}). Three candidate hybrid stars (out of 61 stars with known spectral type), four candidate \dsct\,stars (out of 67 stars), and one candidate \gdor\,star (out of 25 stars) are also Ap or Am stars. So far, we detected both p- and g-mode pulsators among the chemically peculiar stars. 
Balona et al. (2011c) stated that the instability strip of pulsating Am type stars and \dsct\,stars do not differ much. With the current small number statistics, it is not clear whether Ap/Am stars are indeed rare among \gdor\,stars (Handler \& Shobbrook 2002). One of the open questions is if chemical peculiarity is related to hybridity. The first discovered hybrid HD\,8801 (Henry \& Fekel 2005) intruigingly turned out to be an Am star. In a recent abundance study by Hareter et al. (2011) one of the two studied hybrid stars is also confirmed as being a chemically peculiar star. Together with the results of this study,  this brings the total of known chemically peculiar hybrid stars to five. There is currently no evidence for a direct link between chemical peculiarity and hybrid behaviour, but a careful abundance analysis of a  representative sample of hybrid stars is needed to confirm this.

Many more (candidate) \dsct, \gdor, and hybrid stars are expected to be among the stars observed by {\it Kepler}. Debosscher et al. (2011) reported the discovery of many additional \dsct\,and \gdor\,candidates in the public {\it Kepler} Q1 data. Also, a considerable fraction of the host stars of the recently published 1235 {\it Kepler} planet candidates (Borucki et al. 2011)  turn out to be A-F type stars.  Hence, we have promising prospects in studying and understanding the A-F type star variable behaviour in detail through a much larger and more complete sample of A-F stars in the {\it Kepler} field when longer timestrings of {\it Kepler} data will become publicly available.   {\it Kepler} is definitely opening the window towards the accurate characterization of pulsating A-F stars.

\begin{acknowledgements}
We are grateful to Joanna Molenda-\.{Z}akowicz, James Nemec and the anonymous referee for their suggestions and comments to improve this paper. Funding for the {\it Kepler mission } is provided by NASA's Science Mission Directorate. We thank the entire {\it Kepler} team for the development and operations of this outstanding mission. KU acknowledges financial support by the Deutsche Forschungsgemeinschaft (DFG) in the framework of project UY 52/1-1, and by the Spanish National Plan of R\&D for 2010, project AYA2010-17803.  AM acknowledges the funding of AstroMadrid (CAM\,S2009/ESP-1496).  EN and AP acknowledge the financial support of the NN 203 405139 and NN 203 302635 grant, respectively, from the MNiSW. The work by GH and VA was supported by the Austrian Fonds zur F\"orderung der wissenschaftlichen Forschung under grant P20526-N16. LFM acknowledge financial support from the UNAM under grant PAPIIT IN114309. RSz and LLK have been supported by the `Lend\"ulet' program of the Hungarian Academy of Sciences and the Hungarian OTKA grants K83790 and MB08C 81013. RSz was supported by the J\'anos Bolyai Research Scholarship of the
Hungarian Academy of Sciences. SH acknowledges financial support from the Netherlands Organisation for Scientific Research (NWO). This research has been funded by the Spanish grants ESP2007-65475-C02-02, AYA 2010-21161-C02-02 and CSD2006-00070. The research leading to these results has received funding from the
European Community's Seventh Framework Programme (FP7/2007-2013)  under grant agreement no. 269194. This research has made use of the SIMBAD database, operated at CDS, Strasbourg, France, and is partly based on observations obtained at the Observatorio Astr\'onomico Nacional-San Pedro M\'artir (OAN-SPM), Baja California, Mexico, at the Observatoire de Haute Provence, France, and at the Th\"uringer Landessternwarte Tautenburg, Germany. We acknowledge with thanks the variable star observations from the AAVSO International Database contributed by observers worldwide and used in this research.
\end{acknowledgements}

\setcounter{table}{1}
\onltab{1}{
\longtabL{1}{
{\small 
\begin{landscape}

\tablefoot{ 
\centering   
\tablefoottext{1}{P = 9.3562 d, eclipsing binary (Hartman et al. 2004)}                                    
\tablefoottext{2}{P = 0.0665 d (Henry et al. 2001)}                                                        
\tablefoottext{3}{P = 4.0303 d, pulsating star (Hartman et al. 2004)}                                     
\tablefoottext{4}{P = 0.5607 d  (Hartman et al. 2004)}                 \\                                    
\tablefoottext{5}{P = 0.1886 d  (Hartman et al. 2004)}                                                     
\tablefoottext{6}{P = 0.2948 d  (Hartman et al. 2004)}                                                     
\tablefoottext{7}{P = 35.9 d (Watson 2006)}            
\tablefoottext{8}{P =  4.924 d (Pigulski et al. 2009)} 
\tablefoottext{9}{P = 2.18 d (Magalashvili \& Kumishvili 1976)} \\
\tablefoottext{10}{P = 0.066677 d (Watson 2006)}
\tablefoottext{11}{P = 0.38414 d (Watson 2006)}
\tablefoottext{12}{P = 8.4803 d (Otero 2007); 8.480322 d (Carrier et al. 2002)} 
\tablefoottext{13}{P = 4.56427 d, eclipsing binary (Malkov et al. 2006)}\\
\tablefoottext{14}{Abt (1984)}
\tablefoottext{15}{Grigahc{\`e}ne et al. (2010)}
\tablefoottext{16}{Henry et al. (2001)}
\tablefoottext{17}{Sato \& Kuji (1990)}
\tablefoottext{18}{Cannon (1925)}
\tablefoottext{19}{Abt (2004)} 
\tablefoottext{20}{Grenier et al. (1999)}
\tablefoottext{21}{Hill \& Lynas-Gray (1977)} \\
\tablefoottext{22}{Guetter (1968)}
\tablefoottext{23}{Abt \& Cardona (1984)}
\tablefoottext{24}{Perryman et al. (1997)}
\tablefoottext{25}{Macrae (1952)}
\tablefoottext{26}{Floquet (1975)}
\tablefoottext{27}{Floquet (1970)} 
\tablefoottext{28}{Lindoff (1972)} 
\tablefoottext{29}{Moore \& Paddock (1950)} \\
\tablefoottext{30}{Bidelman, Ratcliff \& Svolopoulos (1988)}
\tablefoottext{31}{Stephenson (1986)}
\tablefoottext{32}{Vyssotsky (1958)}
\tablefoottext{33}{Hill \& Schilt (1952)}
\tablefoottext{34}{Cannon (1925)} 
\tablefoottext{35}{Kharchenko \& Roeser (2009)} \\
\tablefoottext{36}{Eclipsing binary (Pr\v{s}a et al. 2011)}
\tablefoottext{37}{Balona et al. (2011c)}
\tablefoottext{38}{Eclipsing binary (Slawson et al. 2011)}
\tablefoottext{39}{Wright et al. (2003)}
\tablefoottext{40}{Hoffleit (1951)}
\tablefoottext{41}{Molenda-\.{Z}akowicz et al. (2008)} \\
$\star$: binarity is suspected by inspection of Digitized Sky Survey and 2MASS images; 
$^{\circ}$: spectroscopic binary
}
\end{landscape}
}}}

\setcounter{table}{2}
\onltab{2}{
\longtabL{2}{
\begin{landscape}
\begin{longtable}{c|ccccc|cccc|cc} 
\caption{Effective temperature (in K), \logg\,(in dex), and \vsini~values (in \kms) for the 750 sample stars}
\\ \hline \hline
& \multicolumn{5}{|c}{\Teff~(K)}&\multicolumn{4}{|c}{\logg~(dex)} & \multicolumn{2}{|c}{\vsini~(\kms)} \\
KIC ID& KIC\tablefootmark{a} & Literature & Literature & Literature &{\bf Adopted$^{\star}$} & KIC\tablefootmark{a} & Literature & Literature &{\bf Adopted$^{\star}$} & Spectra & Spectra\\ 
\hline
\endfirsthead
\caption{continued.}\\ \hline \hline
& \multicolumn{5}{|c}{\Teff~(K)}&\multicolumn{4}{|c}{\logg~(dex)} & \multicolumn{2}{|c}{\vsini~(\kms)} \\
KIC ID& KIC\tablefootmark{a} & Literature & Literature & Literature &{\bf Adopted$^{\star}$} & KIC\tablefootmark{a} & Literature & Literature &{\bf Adopted$^{\star}$} & Spectra & Spectra\\
\hline
\endhead
\hline
\setlength{\tabcolsep}{0.5pt}
01162150          &6870           & \ldots                       & \ldots                       & \ldots                       &{\bf 6870 $\pm$290\tablefootmark{a}}&  3.47& \ldots                        & \ldots                       &{\bf 3.5   $\pm$0.3\tablefootmark{a}}&   \ldots                    &   \ldots                    \\
01294756          &8410           & \ldots                       & \ldots                       & \ldots                       &{\bf 8410 $\pm$290\tablefootmark{a}}&  3.92& \ldots                        & \ldots                       &{\bf 3.9   $\pm$0.3\tablefootmark{a}}&   \ldots                    &   \ldots                    \\
01432149          &7270           & \ldots                       & \ldots                       & \ldots                       &{\bf 7270 $\pm$290\tablefootmark{a}}&  3.99& \ldots                        & \ldots                       &{\bf 4.0   $\pm$0.3\tablefootmark{a}}&   \ldots                    &   \ldots                    \\
01571152$^{\circ}$&7050           & 6500$\pm$130\tablefootmark{b}& 7000$\pm$ 75\tablefootmark{g}& \ldots                       &{\bf 7000 $\pm$ 75\tablefootmark{g}}&  3.16& 4.49$\pm$0.21\tablefootmark{b}&4.15$\pm$0.26\tablefootmark{g}&{\bf 4.2  $\pm$0.26\tablefootmark{g}}& 91$\pm$13\tablefootmark{g}  &   \ldots                    \\
01571717          &8110           & \ldots                       & \ldots                       & \ldots                       &{\bf 8110 $\pm$290\tablefootmark{a}}&  3.98& \ldots                        & \ldots                       &{\bf 4.0   $\pm$0.3\tablefootmark{a}}&   \ldots                    &   \ldots                    \\
\end{longtable}
\tablefoot{
\centering 
$^{\circ}$: spectroscopic binary; 
{\bf Values derived from photometry:} \tablefoottext{a}{KIC Catalogue, Latham et al. (2005)}
\tablefoottext{b}{SPM photometry, this paper}
\tablefoottext{c}{Masana, Jordi,  \& Ribas (2006)} \\
\tablefoottext{d}{Allende Prieto \& Lambert (1999)}
\tablefoottext{e}{Hauck \& Mermilliod (1998)}; 
{\bf Values derived from photometry or spectroscopy:}
\tablefoottext{f}{Balona et al. (2011b)} \\
{\bf Values derived from spectroscopy:}
\tablefoottext{g}{TLS spectra, this paper}
\tablefoottext{h}{SOPHIE spectra, this paper}
\tablefoottext{i}{Catanzaro et al. (2011)}
\tablefoottext{j}{Lehmann et al. (2011)} \\
\tablefoottext{k}{Balona et al. (2011c)}
\tablefoottext{l}{Antoci et al., private communication}
\tablefoottext{m}{Breger et al. (2011)}
\tablefoottext{n}{Glebocki \& Stawikowski (2000)}
\tablefoottext{o}{Nordstr\"om et al. (2004)} \\
\tablefoottext{p}{Catanzaro et al. (2010)}
\tablefoottext{q}{Molenda-\.{Z}akowicz et al. (2008)}
\tablefoottext{r}{Antoci et al. (2011)}; \\ $^{\star}$: the estimated errors  on the KIC values are 290\,K for \Teff\,and 0.3\,dex for  \logg\,(see text).
}
\end{landscape}
}
}

\setcounter{table}{3}
\onltab{3}{
\longtabL{3}{
\begin{landscape}
                                                 
\end{landscape}
}}
                                                
\end{document}